\documentclass[nologo,11pt,a4paper]{ETHpaper}
\usepackage{graphicx, amsmath, amssymb,color,wasysym}
\usepackage{bm}
\usepackage[square,numbers,sort&compress]{natbib}

\begin{document}

\newcommand{\mean}[1]{\left\langle #1 \right\rangle} 
\newcommand{\abs}[1]{\left| #1 \right|} 
\newcommand{\ul}[1]{\underline{#1}}
\renewcommand{\epsilon}{\varepsilon} 
\newcommand{\eps}{\varepsilon} 
\renewcommand*{\=}{{\kern0.1em=\kern0.1em}}
\renewcommand*{\-}{{\kern0.1em-\kern0.1em}} 
\newcommand*{\+}{{\kern0.1em+\kern0.1em}}

\newcommand{\RA}{\Rightarrow}
\newcommand{\bbox}[1]{\mbox{\boldmath $#1$}}

\title{The law of proportionate growth and its siblings: \\ Applications in agent-based modeling of socio-economic systems}

\titlealternative{The law of proportionate growth and its siblings: \\ Applications in agent-based modeling of socio-economic systems}

\author{Frank Schweitzer}
\authoralternative{Frank Schweitzer}

\address{Chair of Systems Design, ETH Zurich, Weinbergstrasse 58, 8092 Zurich, Switzerland}

\reference{(submitted for publication)}

\www{\url{http://www.sg.ethz.ch}}

\makeframing
\maketitle

\begin{center}
  \emph{\large In memory of Masanao Aoki}
\end{center}

\begin{abstract}
  The law of proportionate growth simply states that the time dependent change of a quantity $x$ is proportional to $x$.
  Its applicability to a wide range of dynamic phenomena is based on various assumptions for the proportionality factor, which can be random or deterministic, constant or time dependent.
  Further, the dynamics can be combined with additional additive growth terms, which can be constants, aggregated quantities, or interaction terms.
  This allows to extent the core dynamics into an agent-based modeling framework with vast applications in social and economic systems. 
  The paper adopts this overarching perspective to discuss phenomena as diverse as saturated growth, competition, stochastic growth, investments in random environments, wealth redistribution, opinion dynamics and the wisdom of crowds, reputation dynamics, knowledge growth, and the combination with network dynamics.

  \end{abstract}
\date{\today}

\section{Introduction}

Stochastic systems and their application to economic dynamics have always been at the heart of Masanao Aoki \citep{aoki2011reconstructing}.
His aim was to reconstruct macroeconomic dynamic behaviour from 
of a large collection of interacting agents \citep{aoki2002modeling}, with a particular focus on the dynamics of firms. 
Already early on, Aoki has combined such studies with decentralized optimization problems \citep{aoki1973decentralized}.
What makes his work appealing to me is the clarity, the rigor and the accessibility of his modeling approach.
Linking economic behavior back to generalized stochastic dynamics allows to bridge different scientific disciplines, including applied mathematics and statistical physics.  

Such achievements made Aoki a forerunner in agent-based modeling, the way we want it to be:
Away from mere, and often arbitrary, computer simulations based on ad-hoc assumptions about agent's behavior, towards a formal, tractable and still insightful analysis of interacting systems.  
Current trends in \emph{econophysics} \citep{aoyama2017macro} and \emph{sociophysics} \citep{Schweitzer2018} point into this direction.

In the spirit of Aoki's approch, I will sketch how a \emph{class of multiplicative models} can be fruitfully applied to various dynamic problems in the socio-economic domain.
The core of this model class is the \emph{law of proportionate growth} proposed by R. Gibrat in 1931 to describe the \emph{growth of firms}.
The size of a firm is described by a (positive) variable $x_{i}(t)$.
The law of proportionate growth then states that the growth of a firm, expressed by the time derivative $\dot{x}_{i}(t)$, is proportional to $x_{i}(t)$.
But for the proportionality factor various assumptions can be made. 
In its original version, growth rates have been proxied by random variables, so it is a \emph{stochastic model}. 
It is also an \emph{agent-based model} because the dynamics focuses on individual firms.
But, as in other agent-based models, the aim is \emph{not} to capture the growth of a particular firm in the most precise manner.
Instead, the research interest is in correctly reproducing the \emph{aggregated}, or macro behavior of an ensemble of firms.

It is interesting that, despite its simplicity, the law of proportionate growth is indeed able to reproduce so-called \emph{stylized facts} about the dynamics of firms.
At the same time, it misses one important modeling ingredient, namely \emph{interactions} between agents.
This sets the ground for the following discussions.
We will extend the basic model by introducing \emph{direct} interactions, but also \emph{indirect} interactions via global couplings, for example redistribution mechanisms.
These different interaction terms are combined with different expressions for the \emph{growth term}, which can become also negative.
Further, in addition to the multiplicative growth term, we consider an \emph{additive term} and, for these two terms, combine \emph{stochastic} and \emph{deterministic} dynamics in different ways.

With these assumptions, we obtain various agent-based models that all describe the dynamics of \emph{agents}, via $x_{i}(t)$.
Eventually, we combine this dynamics for the agent variable with another dynamics that changes the \emph{interaction structure} of agents. 
This way, we arrive at a whole ensemble of agent-based models, which all inherit from the same basic dynamics, namely proportional growth, but additional model components allow to capture a plethora of socio-economic phenomena.

\section{Basic Dynamics: Multiplicative Growth}
\label{sec:proportionate-growth}

\subsection{Exponential growth: Short time horizons}
\label{sec:exp}

In the following, we consider a number of agents $i=1,...,N$, each of
which is described by a time dependent quantity $x_{i}(t)$.
Thus, the
system's dynamics results from $N$ concurring dynamic processes.
$x$ is \emph{continuous} and \emph{positive}, i.e. $x_{i}(t)\geq 0$.
The law of proportionate growth states that the increase in time of
$x_{i}(t)$ is proportional to the current value
\begin{equation}
  \label{eq:growth}
  \frac{dx_{i}(t)}{dt}=\alpha_{i} x_{i}(t)\;;\quad
  x_{i}(t)=x_{i}(0)\, \exp\{\alpha_{i}t\}
\end{equation}
It describes a self-reinforcing process which for the growth factor
$\alpha_{i} >0$ leads to exponential growth and for $\alpha_{i}<0$ to
exponential decay of $x_{i}(t)$.
Instead of a continuous time $t$, we can also consider a discrete formulation $t,t+\Delta t, t+2\Delta t, ...$.
With $\Delta t=1$, the growth dynamics then reads:
\begin{equation}
  \label{eq:2}
  x_{i}(t+1)=x_{i}(t)\left[1+\alpha_{i}\right]
\end{equation}
The exponential growth dynamics is shown in Figure~\ref{fig:comp}(a). 

\begin{figure}[htbp]
  \centering
  \includegraphics[width=0.38\textwidth]{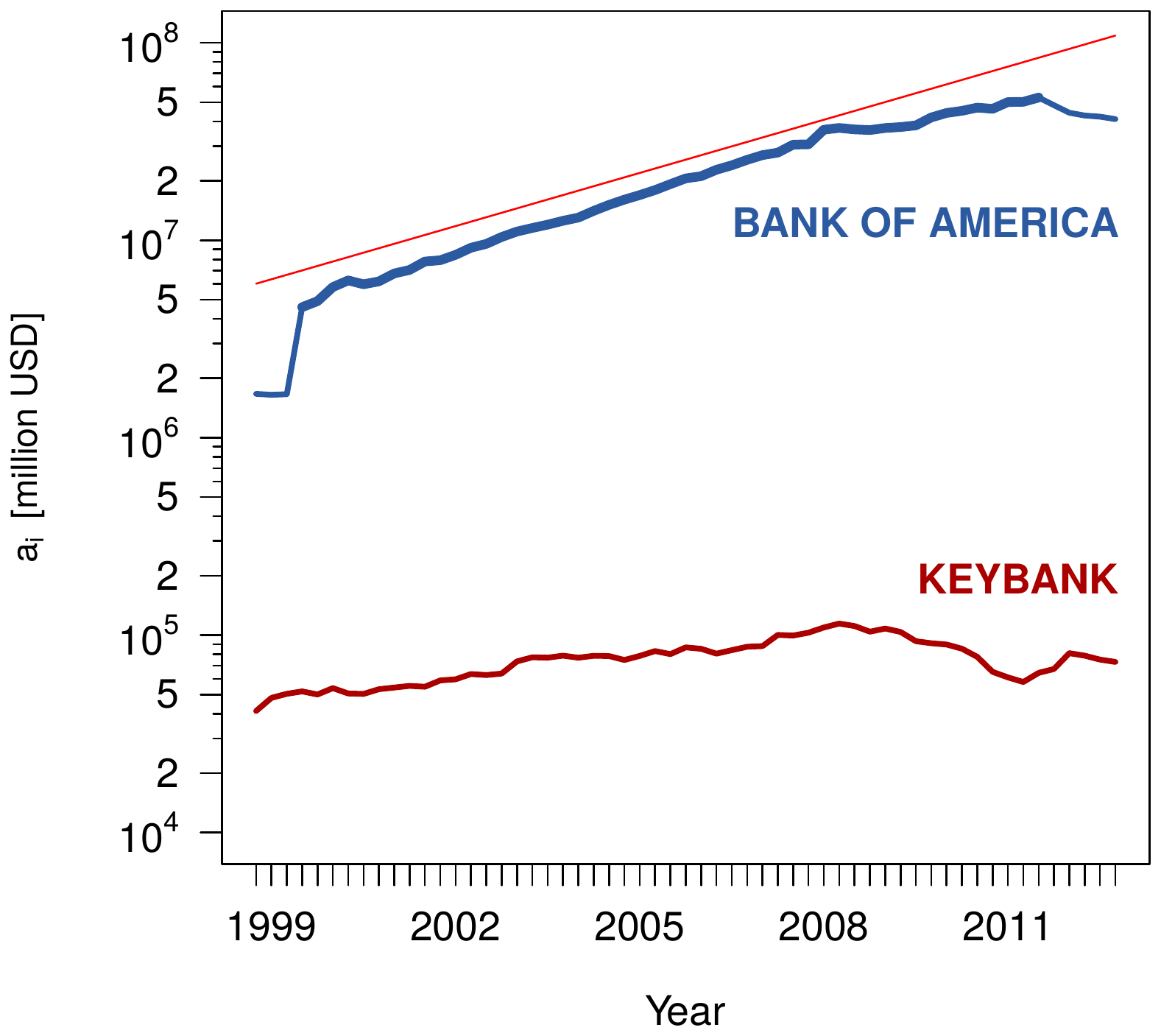}\hfill
  \includegraphics[width=0.45\textwidth]{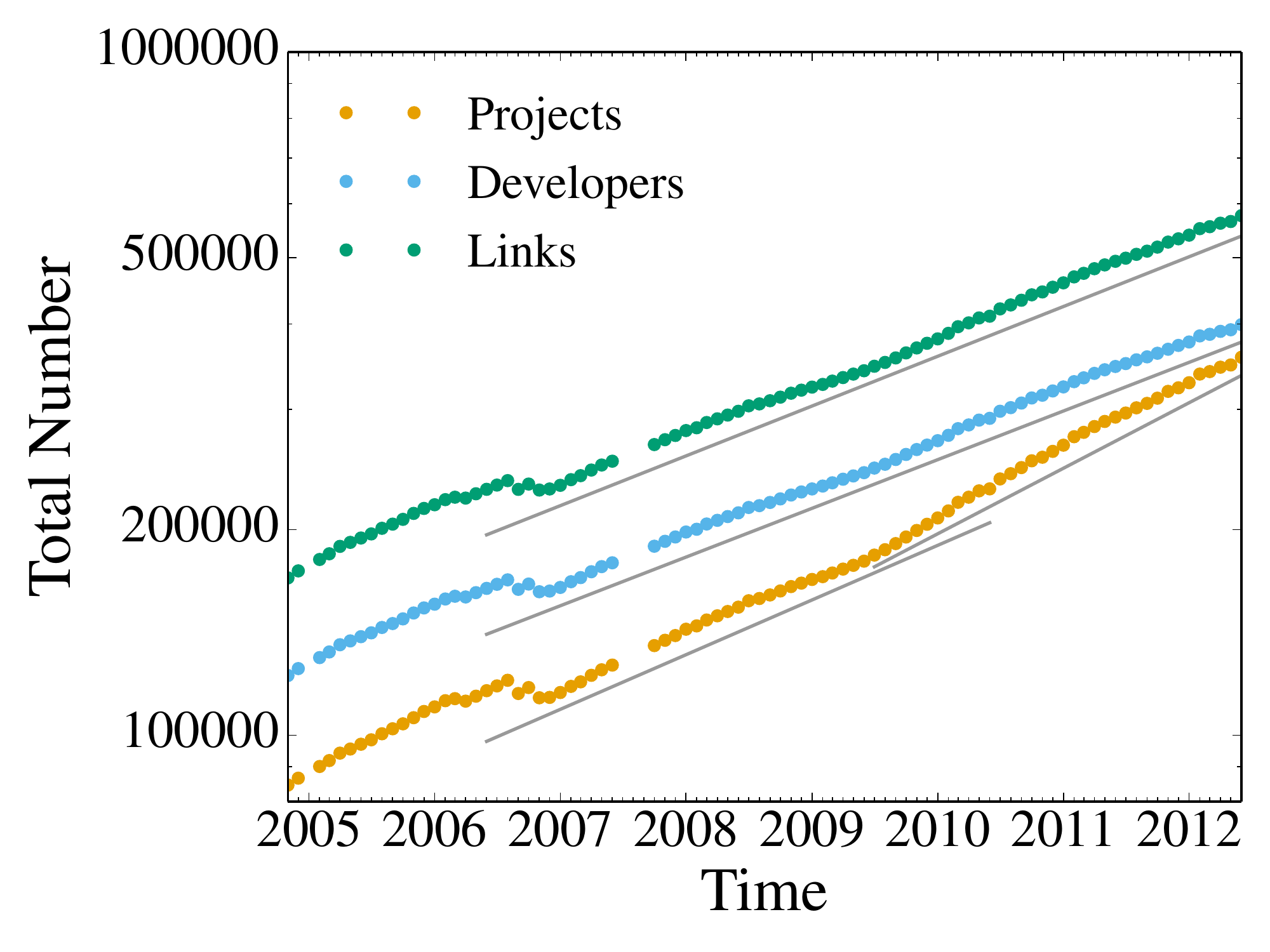}
  \caption{Exponential growth in real systems: (a) Value of OTC derivatives of the Bank of America \citep{Nanumyan2015}, (b) Number of developers registered on the platform \texttt{Sourceforge} \citep{Nanumyan2014} }
  \label{fig:exp}
\end{figure}

\paragraph{Applications: }
Using this dynamics in a plain manner has the problem that, for \emph{long} times, the result becomes either \emph{unrealistic}, because with $\alpha_{i}>0$ the values of $x_{i}$ exceed any limit, or it becomes \emph{uninteresting}, because with $\alpha_{i}<0$ the values of $x_{i}$ converge to zero.
Nevertheless, for \emph{intermediate} times, we can still find exponential growth/decay in real systems. 
Exponential growth has been observed in OTC derivatives market of the US \citep{Nanumyan2015}, but also in the growth of open source software platforms such as  \texttt{Sourceforge} \citep{Nanumyan2014} (see Figure \ref{fig:exp}).  
 
\subsection{Relative growth: Competition}
\label{sec:relat-growth-comp}

\begin{figure}[htpb]
  \centering \includegraphics[width=0.45\textwidth]{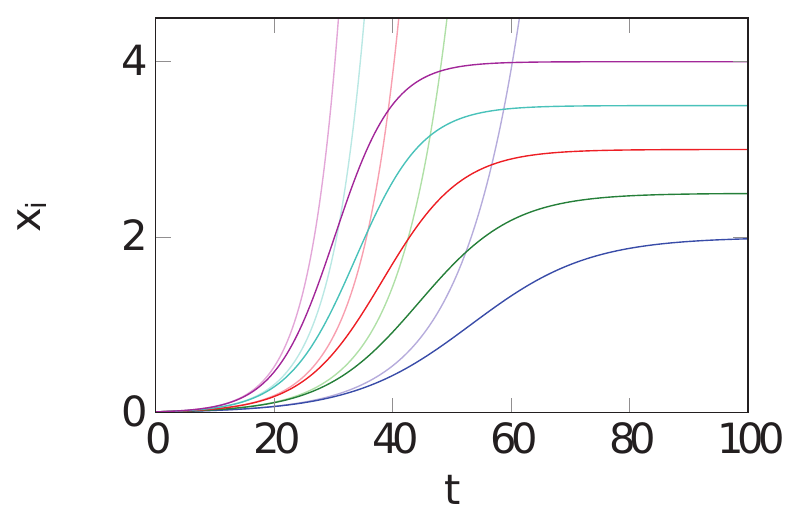}(a) \hfill
  \raisebox{1ex}{\includegraphics[width=0.46\textwidth]{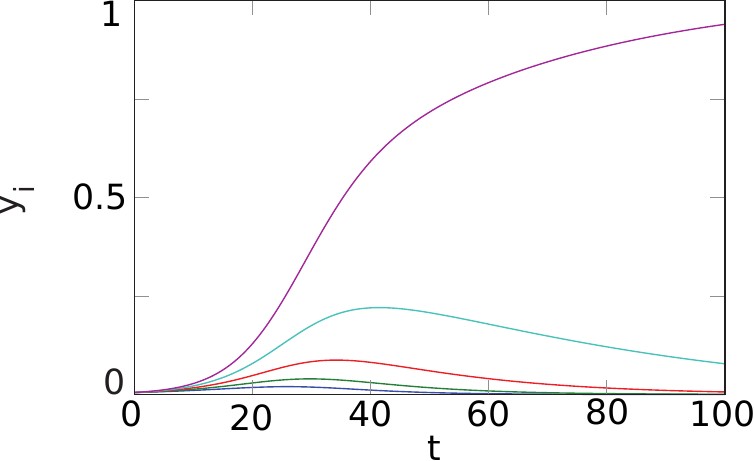}}(b)
  \caption{(a) Exponential growth, Eqn. \eqref{eq:growth}, and saturated growth,
    Eqn.~\eqref{eq:xsat}. (b) Relative growth, Eqn.~\eqref{eq:rel}. Parameters: $a_i$ (0.1; 0.125;
    0.15; 0.175; 0.2), $b_{i}$=0, saturated growth:
    $b_{i}$=0.05. }
  \label{fig:comp}
\end{figure}

If we consider concurring processes, it is useful to introduce relative
variables, or fractions, $y_{i}(t)=x_{i}(t)/\sum_{i}x_{i}(t)$ for which
a conservation holds: $\sum_{i} y_{i}(t)=1$. Let us further
assume some direct or indirect coupling that may
affect the growth rate $\alpha_{i}=\alpha_{i}(...., x_{i},x_{j},...)$.
For example, the growth of quantity $x_{i}$ occurs via an
interaction between agents $i$ and $j$. Specificall it depends on the
\emph{relative advantage} $(a_{i}-a_{j})$ of $i$ over $j$ and the
respective quantity $x_{j}(t)$. If all agents are allowed to interact,
for the growth of $i$ results:
\begin{equation}
  \label{eq:axij}
  \alpha_{i}(...., x_{i},x_{j},...) = \sum\nolimits_{i}
\left(a_{i}-a_{j}\right) x_{j}
\end{equation}
For the growth dynamics of agent $i$ we then find, in terms of the
relative variable:
\begin{equation}
  \label{eq:rel}
  \frac{dy_{i}(t)}{dt}=y_{i}(t)\left[a_{i}-\mean{a(t)}\right]\;;
\quad \mean{a(t)} = \frac{\sum_{i}a_{i} x_{i}(t)}{\sum_{i }x_{i}}
=\sum\nolimits_{i}a_{i}    y_{i}(t)
\end{equation}
This selection equation, also known as Fisher-Eigen equation \citep{feistel2011physics}, states that,
despite the $x_{i}(t)$ of all agents grow exponentially, their share relative to the
total population only grows as long as their advantage (or fitness)
$a_{i}$ is larger than the average fitness $\mean{a(t)}$. The latter,
however, increases over time because the $x_{i}(t)$ of agents with a fitness \emph{below average}
shrink. This way, each agent $i$ receives
indirect information about its performance with respect to the average
performance.
In the end, this competition dynamics leads to an outcome where
only one agent survives -- the one with the highest advantage $a_{i}$.
This is illustrated in Fig. \ref{fig:comp}(b).

\paragraph{Applications: }
The competition scenario described by Eqn. (\ref{eq:rel}) holds if (i) a
self-reinforcing growth mechanism is involved, and (ii) a conservation
law applies.
This is quite common in socio-economic systems, where e.g. products compete for customers via their cost price \citep{feistel2011physics}.
Also, market shares cannot grow independently because they are coupled to the market size \citep{fs-98-jcs}.
The same dynamics can be also found for competing strategies in a game-theoretical setting \citep{Mavrodiev2013} or in cluster formation \citep{fs-lsg-94}.

\subsection{Size dependent growth factor: Saturation}
\label{sec:size-depend-growth}

Long-term exponential growth is a quite unrealistic scenario if limited
resources are considered. Therefore, for the growth factor $\alpha_{i}$
usually some quantity dependent decrease is assumed. A very common
assumption is
\begin{equation}
  \label{eq:sat}
  \alpha_{i}(x_{i})= a_{i} -b_{i}x_{i}
\end{equation}
where $b_{i}$ is assumed to be small. In this case, we observe saturated
growth (see also Fig. \ref{fig:comp}a)
\begin{equation}
  \label{eq:xsat}
  \frac{dx_{i}(t)}{dt}=a_{i} x_{i}-b_{i}x_{i}^{2}\;;\quad x_{i}(t\to
  \infty)
=\frac{a_{i}}{b_{i}}
\end{equation}
which is known as the \emph{logistic equation}, originally proposed by P. Verhulst in 1838, rediscovered by R. Pearl in 1920 and by A. Lotka in 1925.  

\paragraph{Applications: }
In \emph{population dynamics} many realistic growth processes are described by the logistic equation, where the saturation reflects a limited \emph{carrying capacity}. 
Surprisingly, also the growth of \emph{donations} empirically matches this dynamics \cite{Schweitzer2008}. Here, the limited resource is not only the money available, but also the number of people willing to donate.  

An important application of the dynamics of Eqn.~\eqref{eq:xsat} comes from its  discretized version.
Using the transformation $t\to n$, $z_{n}=x(t)\ b/r$ with $r=a+1$, we arrive at the famous \emph{logistic map}
\begin{equation}
  \label{eq:3}
  z_{n+1}= r\ z_{n}(1-z_{n})
\end{equation}
This is one of the paragons to study \emph{deterministic chaos}, provided that $3.57<r<4$.

\subsection{Time dependent growth factor: Randomness}
\label{sec:rand}

\begin{figure}[htpb]
  \centering
\includegraphics[width=0.45\textwidth]{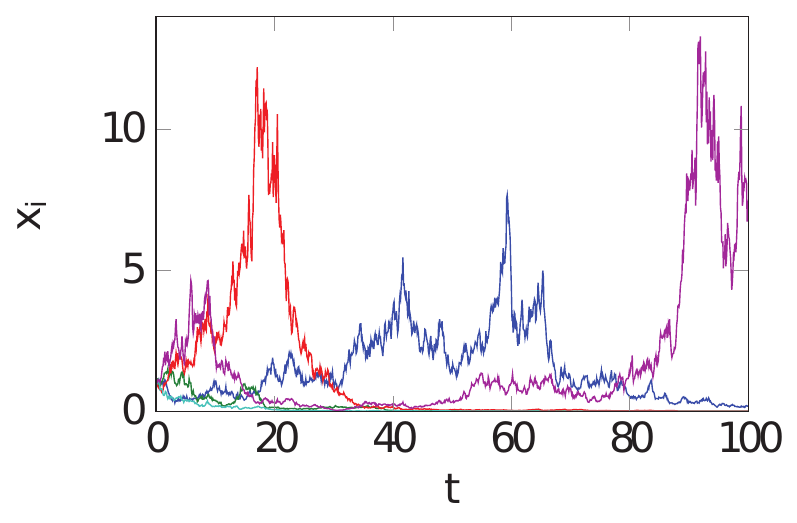}(a) \hfill
\includegraphics[width=0.45\textwidth]{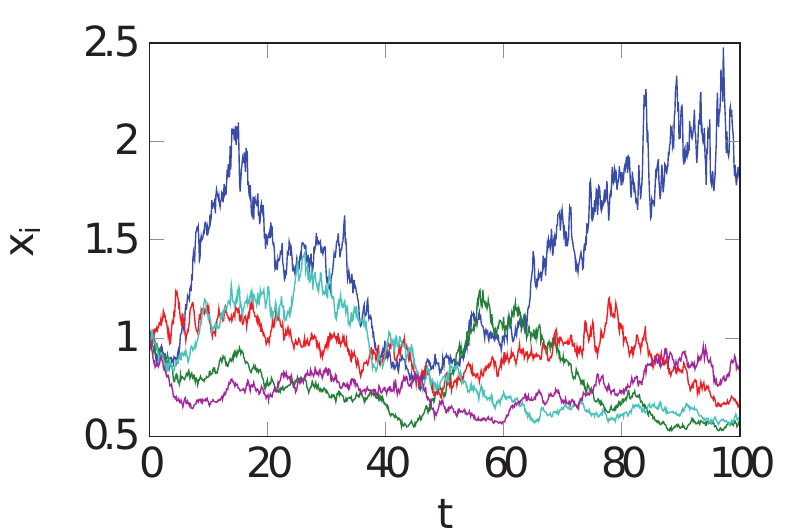}(b)
\caption{(a) Random growth, Eqn.~\eqref{eq:rand}, (b) Coupled random growth, Eqn.~\eqref{eq:polya}. Parameters: $\mathcal{N}(\mu_{\eta},\sigma^{2}_{\eta})$ with $\mu_{\eta}$=0, $\sigma^{2}_{\eta}$=1.
  }
  \label{fig:random}
\end{figure}

An important variant of the law of proportionate growth assumes random
growth factors instead of fixed ones, i.e $\alpha_{i} \to \eta_{i}(t)$, where $\eta_{i}(t)$ is a random number draw from e.g. a normal distribution
with mean $\mu_{\eta}$ and variance $\sigma^{2}_{\eta}$: 
\begin{equation}
  \label{eq:rand}
\eta_{i}(t)\sim \mathcal{N}(\mu_{\eta},\sigma^{2}_{\eta})
\end{equation}
 Because of
stochastic influences the most successful agent cannot be predicted from
the outset, as shown in Fig. \ref{fig:random}(a). Instead, one finds that
the quantity $x$ follows a log-normal distribution $P(x,t)$ which changes
over time as
\begin{equation}
  \label{eq:lognormal}
    P(x,t)=\frac{1}{\sqrt{2\pi
\sigma_{\eta}^{2}t}}\ \frac{1}{x}\  \exp\left\{-\frac{\left[\log(x)-\mu_{\eta}
        t\right]^{2}}{2\sigma_{\eta}^{2}t}\right\}
\end{equation}
We note that this probability distribution does not reach a steady state.
Its mean value and variance increase over time as $\mu_{x}\propto t$, $\sigma^{2}_{x}\propto t$. However, if $\mu_{\eta}<0$, for long times and
sufficiently large $x$ the tail of the distribution can be approximated
by a power law: $P(x)\propto x^{-\abs{\mu_{\eta}}}$.

\paragraph{Applications:} Historically, this dynamics was first used by R. Gibrat in 1931 to describe the growth of companies \citep{sutton1997gibrat}.
This has found mixed empirical evidence \citep{aoyama2017macro}.
But other quantities that can be approximately described by a time-dependent log-normal distribution,  for instance the wealth distribution, have been also modeled with this approach \citep{yakovenko2009colloquium}. 
Eventually, also the growth of cities has been described by the law of proportionate growth \citep{malevergne2011testing}. 

\subsection{Coupled random growth: Fraction dependent fluctuations}
\label{sec:polya}

The dynamics $\dot{x}_{i}(t)=\eta_{i}(t)x_{i}(t)$ with a randomly drawn growth rate $\eta_{i}$ assumes that agents are subject to fluctuations in the same manner, regardless of their value $x_{i}(t)$.
If $x_{i}$ denotes for example the size of a firm, then it is empirically known that \emph{larger} firms should be subject to \emph{smaller} fluctuations \citep{aoyama2017macro}.
This can be considered by a size dependent variance, $\sigma^{2}(x)\propto x^{-\beta}$, where $\beta\approx 0.2$.

Focusing on the size only, however, completely ignores the market structure.
Firms with a larger market share, for instance, face a stronger competition, which should result in larger fluctuations.
One way of achieving this is a global coupling via $\sum_{i}
x_{i}(t)$, for example:
\begin{equation}
  \label{eq:polya}
  \frac{dx_{i}(t)}{dt}=\eta_{i}(t) \frac{x_{i}(t)}{\sum_{i}x_{i}(t)}
=\eta_{i}(t)y_{i}(t)
\end{equation}
which is illustrated in Figure~\ref{fig:random}(b). 
Here, the growth is proportional to the relative influence of agent $i$,
i.e. to \emph{fraction} obtained in the system. For sufficiently chosen
$\mu_{\eta}$, $\sigma^{2}_{\eta}$, the total quantity $\sum_{i}
x_{i}(t)$ may grow over time, which results in a smaller and smaller
impact of the further growth \emph{if} the market share of $i$ is small.
I.e., in the course of time, for those agents we observe a comparably stable value of $y_{i}(t)$.
For agents with a large $x_{i}(t)$, and hence a large fraction $y_{i}(t)$, we still observe remarkable fluctuations, although not comparable to those without a global coupling, as shown in Figure~\ref{fig:random}(a).

In comparison to Eqn. (\ref{eq:axij}), which already introduced such a global coupling between the different
growth processes, the existence of large fluctuations prevent the system from converging to an equilibrium state where ``the winner takes it all''. 
We note that both dynamics of Eqs.~\eqref{eq:rel}, \eqref{eq:polya} belong to the class of so-called \emph{frequency dependent processes}, where the dynamics depends on the relative share $y_{i}$, determined either in a local neighborhood or globally.
Examples of important frequency-dependent processes are the (non-linear) \emph{voter model} \citep{Schweitzer2009d} and the Polya process.

\paragraph{Applications. }

This dynamics allows to combine two processes: (a) the indirect interaction via an evolving mean value, which is the essence of a competition process, and (b) fluctuations in the growth process that depend on the relative influence, or the \emph{ranking}, of the agent.
This combination prevents the system from converging to an uninteresting equilibrium state, but still considers the ``comparative advantage'' of agents.

\section{Multiplicative and additive growth}
\label{sec:mult-addit-growth}

\subsection{Lossy multiplicative growth: Geometric versus arithmetic mean}
\label{sec:lossy-growth}

The outcome of a discrete dynamics of the type
\begin{equation}
  {x}_{i}(t+1)=x_{i}(t)[1+\eta_{i}(t)] = \lambda_{i}(t)x_{i}(t)
  \label{eq:7a}
\end{equation}
very much depends on the \emph{parameters} of the distribution $\mathcal{N}(\mu_{\lambda},\sigma^{2}_{\lambda})$ of the randomly drawn growth rates $\lambda_{i}$.
We can express these parameters as follows:
\begin{equation}
  \label{eq:6}
  \mu_{\lambda}=\mean{\lambda} \;;\quad \sigma^{2}_{\lambda}=\mean{\lambda^{2}}-\mean{\lambda}^{2}
\end{equation}
Equation~\eqref{eq:7a} can be rewritten as
\begin{equation}
  \label{eq:10}
  \log x_{i}(t+1)=\log \lambda_{i}(t) + \log x_{i}(t)=\sum_{s=0}^{t} \log \lambda_{i}(s)
\end{equation}
with the parameters for the distribution of the random variable $(\log \lambda)$:
\begin{equation}
  \label{eq:7}
  \mu_{\log \lambda}=\mean{\log \lambda} \;;\quad \sigma^{2}_{\log \lambda}=\mean{\log \lambda^{2}}-\mean{\log \lambda}^{2}
\end{equation}
Applying the central limit theorem to Eqn.~\eqref{eq:10} implies that the distribution of the random variable $x(t)$ over time gets closer to a log-normal distribution, Eqn.~\eqref{eq:lognormal} or, equivalently, the random variable 
$\log x(t)$ gets closer to a normal distribution with the parameters
\begin{equation}
  \label{eq:11}
  \mu_{\log x}(t)=t\ \mu_{\log \lambda} \;; \quad \sigma_{\log x}^{2}(t)= t \ \sigma^{2}_{\log \lambda}. 
\end{equation}
This means that the \emph{expected value}, i.e. the maximum of the probability distribution, still grows in time.
On the other hand, one can show \cite{Lorenz2013} that individual growth trajectories $x_{i}$ disappear if
\begin{equation}
  \label{eq:8}
  \mu_{\log \lambda}<0< \log \mu_{\lambda} \quad \Rightarrow \quad \mean{\lambda}_{\mathrm{geo}} =\exp(\mu_{\log \lambda}) <1<\mu_{\lambda}=\mean{\lambda}
\end{equation}
where $\mean{\lambda}_{\mathrm{geo}}$ denotes the \emph{geometrical mean}, which has to be smaller than the \emph{arithmetic mean} $\mean{\lambda}$.
The fact that $x_{i}(t)\to 0$ is remarkable because it is also counter intuitive.
So, there is a need to also use agent-based modeling in addition to an analysis of aggregated measures, such as distributions, because it allows us to better understand what happens on the microscopic/individual level.

\paragraph{Applications. }

The above insights can be directly applied to multiplicative growth processes with random time-dependent growth factors discussed already in Sect.~\ref{sec:rand}.
Hence, they help to better understand under which conditions a decline in firm sizes, city sizes, or individual wealth \citep{slanina2004inelastically} can be expected if the underlying stochastic dynamics holds. 
The same dynamics was also applied to model the growth, more precisely the decline, of individual human capital which follows a life cycle over time \citep{grochulski2010risky}.  

\subsection{Constant additive growth: Stationarity}
\label{sec:const-addit-growth}

To avoid a scenario where individual growth trajectories disappear, one can add a term $\omega_{i}$ to the dynamics:
\begin{equation}
  \label{eq:1}
x_{i}(t+1)=\lambda_{i}(t)x_{i}(t) + \omega_{i}(t)
\end{equation}
$\omega_{i}(t)$ can have different forms as discussed below: it can be a small positive constant, $\omega_{i}(t)\equiv A >0$, it can be a time dependent function that further depend on the state of other agents, or it can be fluctuating, like an additive noise term.

\begin{figure}[htbp]
  \centering
 \includegraphics[width=0.45\textwidth]{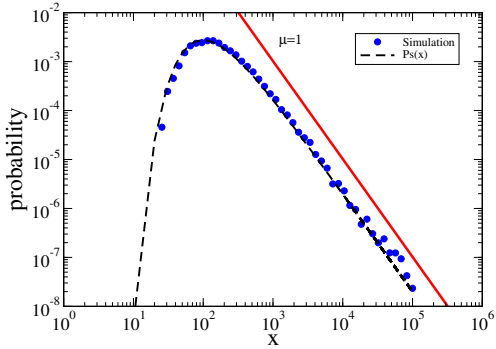}(a)
 \hfill
 \raisebox{2ex}{\includegraphics[width=0.45\textwidth]{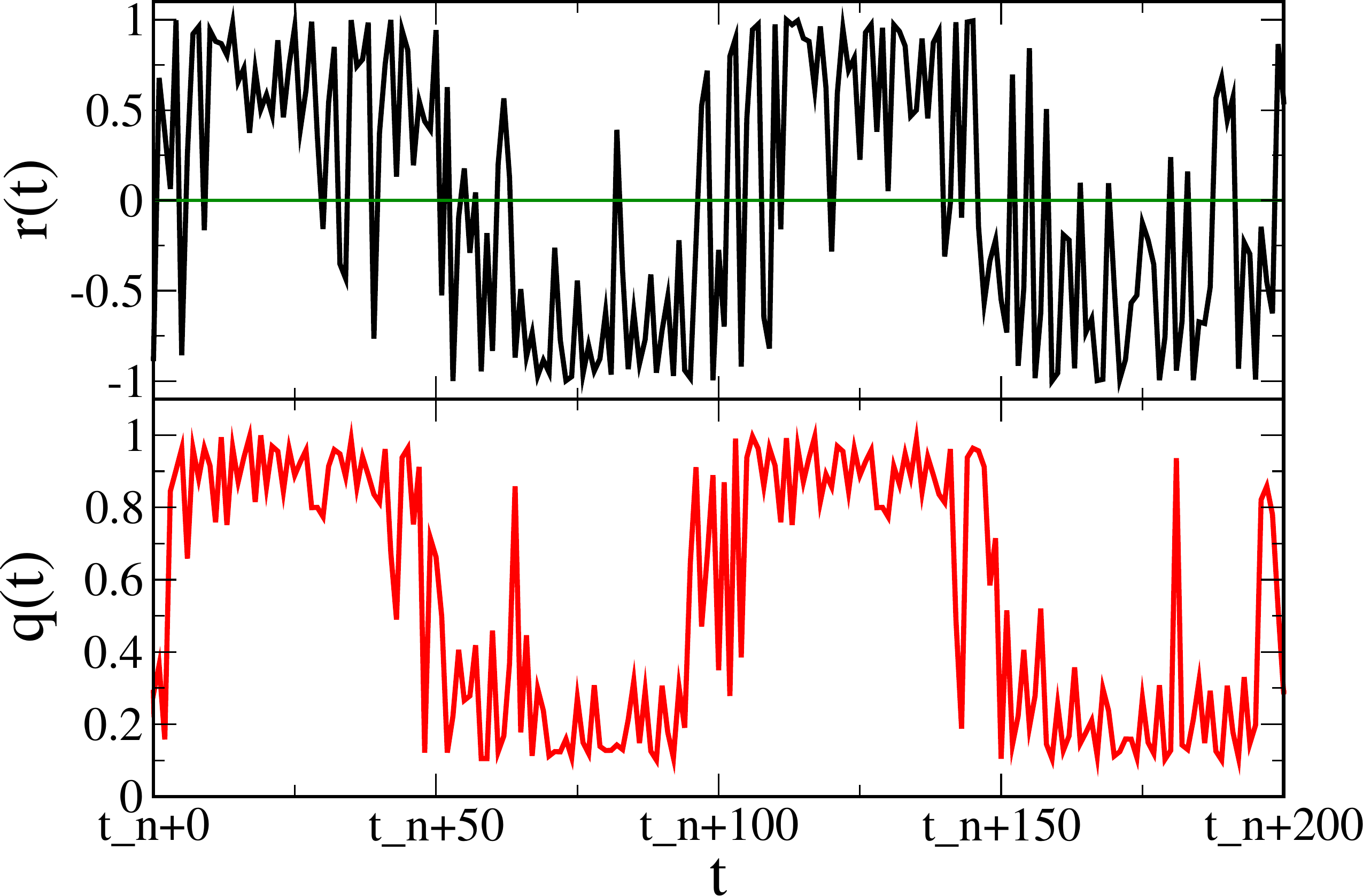}}(b)
 \caption{(a) Stationary distribution $P^{s}(x)$, Eqn.~\eqref{eq:9} (dashed line), agent-based simulations, Eqn. \eqref{eq:1} (dots) \citep{Navarro2008}, (b) Detecting an optimal investment $q_{i}(t)$ in a noisy market by means of a genetic algorithm \citep{NAVARRO-BARRIENTOS2008}. }
  \label{fig:stat}
\end{figure}

The mere existence of such an additive term changes the properties of the underlying dynamics.
For $\omega_{i}=A$, we find a \emph{stationary} distribution:
\begin{equation}
  \label{eq:9}
  P^{s}(x)=\frac{(A/\sigma^{2}_{\lambda})^{\mu_{\lambda}}}{\Gamma(\mu_{\lambda})} x^{-(1-\mu_{\lambda})}
  \exp\left\{-\frac{A}{\sigma^{2}_{\lambda} x}\right\}
\end{equation}
where $\Gamma(x)$ is the Gamma function.
This distribution is plotted in Figure \ref{fig:stat}.
The most probable value, i.e. the maximum of the distribution, is given by $x^{\mathrm{mp}}\approx A/\mean{\lambda^{2}}$.

We note that Eqs.\eqref{eq:1}, \eqref{eq:9} are special cases of a more general framework for multiplicative processes \citep{richmond2001power}
\begin{equation}
  \label{eq:4}
  \Delta x(t)=\eta(t) G[x(t)] + F[x(t)]
\end{equation}
with the general (non-normalized) solution
\begin{equation}
  \label{eq:5}
  P^{s}(x)=\frac{1}{G^{2}(x)} \exp\left\{\frac{2}{D} \int^{x} \frac{F(x^{\prime})}{G^{2}(x^{\prime})} \mathrm{d}x^{\prime} \right\}
\end{equation}
Our case is covered by $G(x)=x$ and $F(x)=A$.

\paragraph{Applications. }
This dynamics is frequently used to model stock market behavior \citep{richmond2001power} or investments in random environments, in general \citep{Navarro2008,NAVARRO-BARRIENTOS2008}.
The stochasticity can come from fluctuating yields, e.g. $\eta_{i}(t)=r(t)q_{i}(t)$,
where $0\leq q_{i}(t)\leq 1$ is the wealth fraction that agent $i$ decides to invest in a volatile market and  $r(t)$ is the \emph{return on investment}, i.e. the random variable.
$r(t)$ is independent of the agents and describes the market dynamics, with a lower bound of $-1$, i.e. full loss of the investment, but no upper bound.
Investment decisions are then modeled by forecasting the best value $q_{i}(t)$ given some information about previous values of $r(t)$ from time series data.
Here, machine learning algorithms can be used to determine the dynamics for $q_{i}(t)$, as demonstrated in Figure~\ref{fig:stat}(b).

\subsection{Variable additive growth: Redistribution}
\label{sec:vari-addit-growth}

Instead of a fixed (small) amount added to the individual growth dynamics, one can also consider a changing amount that depends on the overall growth.
Let us assume that $x_{i}(t)$ denotes the individual wealth of agent $i$, which is taxed by a a central authority (the government) at a fixed tax rate
$a$, known as \emph{proportional tax} scenario.

From the total amount of taxes, $T(t)= a \sum_{i} x_{i}(t)$, the government withholds a fraction $b$ to cover the costs for its administration, and \emph{redistributes} the remaining fraction  $(1-b)$ equally to all $N$ agents as a subsidy \citep{Lorenz2013}.
The wealth of each agent still evolves independently according to the stochastic growth dynamics of Eqn.~\eqref{eq:7}.
Together with the taxation and the subsidy, the total wealth of an agent at time $t+1$ is therefore given as: 
\begin{equation}
  \label{eq:wi}
  x_{i}(t+1)= x_{i}(t)\left[\lambda_{i}(t) - a \right] + \frac{a[1-b]}{N} \sum_{i}  x_{i}(t)
\end{equation}
Let us now assume that the conditions of Eqn. \eqref{eq:8} hold, i.e. that due to the stochastic dynamics alone the individual wealth will disappear over time.
This is realized by choosing $\mean{\lambda}_{\mathrm{geo}}=2/3<1<3/2=\mean{\lambda}$.
The larger the spread of these two values, the larger the ``risk'' associated with the ``production of wealth''. 
The question then is: under which conditions could the proposed redistribution mechanism prevent the decay of wealth?
Could it even lead to an \emph{increase}, instead of a decrease, of individual wealth over time? 

\begin{figure}[htbp]
  \begin{minipage}[c]{.53\linewidth}
  \centering
    \includegraphics[width=0.48\textwidth]{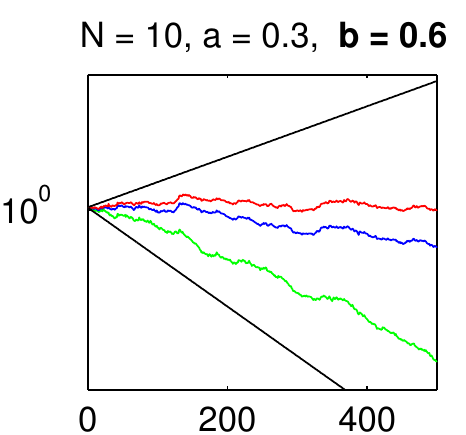} \hfill
    \includegraphics[width=0.48\textwidth]{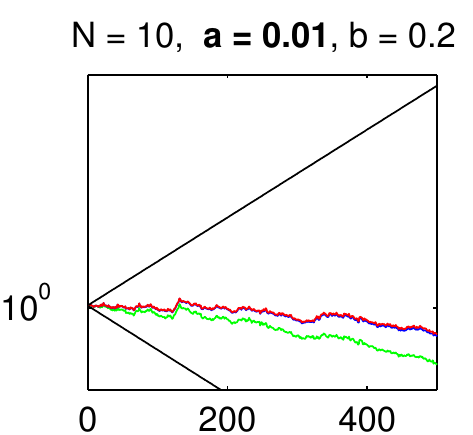}\\
    \includegraphics[width=0.48\textwidth]{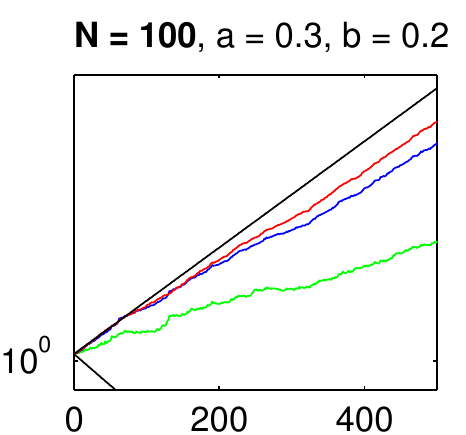}(a)
  \end{minipage}
  \begin{minipage}[c]{.43\linewidth}
   \centering \includegraphics[width=0.99\textwidth]{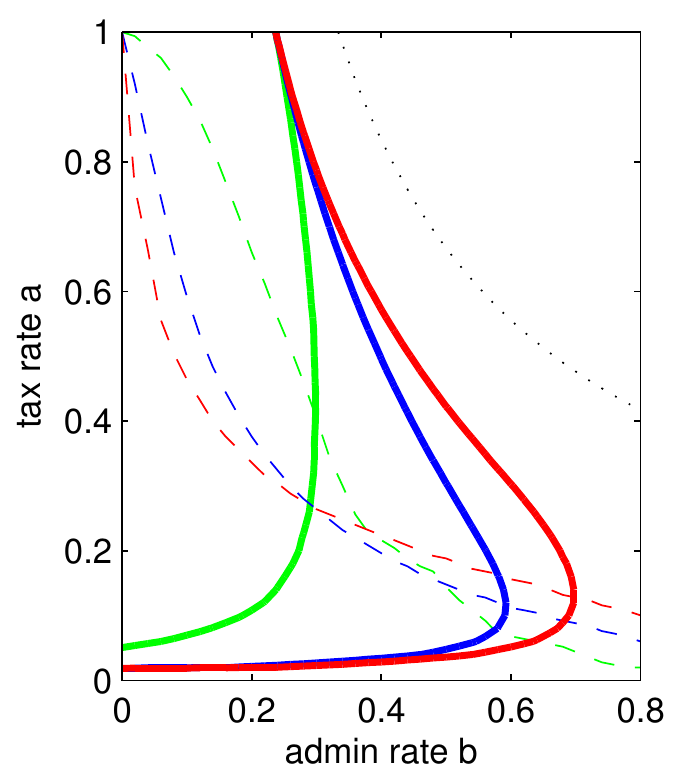}(b)  
  \end{minipage}
  \caption{(a) Sample trajectories of the {total wealth} $\sum_{i}x_{i}(t)$ ($y$-axis, in $\log$ scale) over time $t$ ($x$-axis, in normal scale) for three different parameter settings $a$, $b$, $N$. Tax schemes: (blue) proportional tax, (red) progressive tax (no tax if wealth is below a threshold), (green) regressive tax (fixed tax for everyone), see \citep{Lorenz2013} for details.
  (b) Solid lines divide zones of wealth growth (to the left) and wealth destruction (to the right). Dashed lines $a_{\mathrm{opt}}(b)$ are optimal tax rates for a given administration cost, which maximize the growth of total wealth. Above the black dotted line only wealth destruction can happen.}
  \label{fig:traj}
\end{figure}

The answer is \emph{yes}, and the simplicity of the agent-based model allows to study these conditions in a simulation approach.
Figure~\ref{fig:traj}(a) shows sample trajectories of the \emph{total wealth} $\sum_{i}x_{i}(t)$ for varying parameters $a$, $b$, $N$.  The colors refer to three different taxation schemes (see \citep{Lorenz2013} for details), the blue curve holds for proportional tax.
The straight black lines show two limit cases: The lower line is for ``no tax'', where the individual wealth evolves in time proportional to $[\mean{\lambda}_{\mathrm{geo}}]^{t}$, i.e. it decays exponentially.
The upper line is for ``full tax'', where the individual wealth evolves over time proportional to $[(1-ba)\mean{\lambda}]^{t}$, i.e. it increases exponentially.
Realistic redistribution scenarios have to be between these two limit cases and further depend on the size of the agent population, $N$.
The larger the population, the better the redistribution effect.
Obviously, the value of the tax rates, $a$, is not to be chosen independently of the value of the administration cost, $b$.
Figure~\ref{fig:traj}(b) shows, for fixed values of $N$ and $\mean{\lambda}$, $\mean{\lambda}_{\mathrm{geo}}$, the range of parameters which could possibly lead to an increase of total wealth.

\paragraph{Applications. } 

The redistribution model allows to study the impact of different \emph{tax scenarios} on the wealth of a population.
As shown in Figure~\ref{fig:traj}, two other realistic scenarios of tax collection, progressive tax and regressive tax, have been discussed in \citep{Lorenz2013}.
Further, the impact of different \emph{redistribution mechanisms} can be studied and additional economic constraints, such as conservation of money, can be included.

From a more general perspective this redistribution model has much in common with other models studying the \emph{portfolio effect} in \emph{investment} science \citep{marsili1998dynamical,slanina1999possibility}.
The positive impact of rebalancing gains and losses, first discussed by JL Kelly in 1956, is rediscovered from time to time in different contexts \citep{yaari2010cooperation}.

\section{Multiplicative decay and additive growth}
\label{sec:expon-decay-addit}

\subsection{Additive stochastic influences: Brownian agents}
\label{sec:addit-stoch-infl}

So far, we assumed that the proportional growth term $\alpha_{i}x_{i}(t)$ has a positive growth rate $\alpha_{i}$ at least \emph{on average}. 
Otherwise, instead of growth, we can only observe an exponential decay of $x_{i}$ over time, which needs to be compensated by additional \emph{additive terms}.
There is a whole class of dynamic processes where $\alpha_{i}$ is \emph{always} negative.
For example, the motion of a particle under friction is described by a friction coefficient $\alpha_{i}\equiv - \gamma$.
For the case of \emph{Brownian particles}, the equation of motion proposed by Langevin posits that this friction is compensated by an \emph{additive stochastic force}, to keep the particle moving:
\begin{align}
  \label{eq:1a}
  \frac{d v(t)}{d t} &= - \gamma \, v(t) + \sqrt{2S}\ {\xi (t)} \;; \quad
                        \mean{\xi(t)}=0 \;; \quad \mean{\xi(t')\xi(t)}=\delta(t' -t) 
\end{align}
$S$ denotes the strength of the stochastic force and is, in physics, determined by the  \emph{fluctuation-dissipation theorem}. 
$\xi(t)$ is Gaussian white noise, i.e. it has the expectation value of zero and only delta-correlations in time. 
$v(t)$ denotes the continuous \emph{velocity} of the Brownian particle, which can be positive or negative.
The positive quantity $x_{i}(t)=\abs{v_{i}(t)}$ then has the physical meaning of a speed and follows the same equation.  

We assume that the \emph{agent dynamics} is described by a set of stochastic equations which resemble the Langevin equation of Browian motion, therefore the notion of \emph{Brownian agents} has been established \citep{schweitzer2007brownian}.
For our further consideration in an \emph{agent-based model}, the \emph{structure} of Eqn. \eqref{eq:1a} is important.
The agent dynamics results from a \emph{superposition} of two different types of influences, deterministic and stochastic ones. 
In Eqn. \eqref{eq:1}, the first term $\lambda_{i}(t)x_{i}(t)$ is the stochastic term, while the second term $\omega_{i}(t)$ is the deterministic term.
In Eqn. \eqref{eq:1a}, on the other hand, the first term denotes \emph{deterministic} forces.
This is, in the most simple case of Eqn. \eqref{eq:1a}, the relaxation term which defines a temporal scale of the agent dynamics. 
The second term denotes \emph{stochastic} forces which summarize all  influences that are \emph{not} specified on these temporal or spatial scales.

To develop Eqn. \eqref{eq:1a} into the dynamics of a Brownian \emph{agent}, this picture still misses \emph{interactions} between agents. These can be represented by additional additive terms:
\begin{align}
  \frac{d x_{i}(t)}{d t} &= - \gamma \, x_{i}(t) + \mathcal{G}(\mathbf{x},\mathbf{w},\mathbf{u}) + D_{i}{\xi_{i}(t)} 
  \label{eq:b}
\end{align}
The function $\mathcal{G}(\mathbf{x},\mathbf{w},\mathbf{u})$ fulfills several purposes.
First, with $\mathbf{x}$ as the vector of all variables $x_{i}(t)$, it describes interactions between agents via couplings between $x_{i}(t)$ and any $x_{j}(t)$.
Second, $\mathcal{G}$ is, in general, a \emph{nonlinear} function of $x_{i}$ itself, $\sum_{k=0}^n \beta_k(\mathbf{w},\mathbf{u}) \ x_{i}^{k}(t)$ \citep{Schweitzer2018}, which allows to consider dynamic feedback processes such as self-reinforced growth.
Third, the coefficients $\beta_k(\mathbf{w},\mathbf{u})$ of such a non-linear function can consider additional couplings to other variables $w_{i}(t)$ of the agents, which are summarized in the vector $\mathbf{w}$.
$\mathbf{u}$ eventually represents a set of \emph{control parameters} to capture e.g. the influence of the environment. 
$D_{i}$ defines the individual susceptibility of agent $i$ to stochastic influences.

\paragraph{Applications. }

The concept of \emph{Brownian agents} \citep{schweitzer2007brownian} has found a vast range of applications at different levels of organization, physical, biological and social. 
Specifically, active motion and clustering in biological systems \citep{fs-lsg-94,Ebeling1999,Garcia2011d,Ebeling2003}, self-wiring of networks and trail formation based on chemotactic interactions \citep{helbing-fs-et-97,Schweitzer.Lao.ea1997Activerandomwalkers,fs-tilch-02-a} and emotional influence in online communications \citep{Schweitzer2010a,Garas2012,Schweitzer2014a,Schweitzer2016a,Tadic2017} are studied both from a modeling and a \emph{data-driven} perspective.

An important application for the additional coupling between agent variables is to consider the variable $w_{i}(t)$ as the \emph{internal energy depot} of an agent.
It allows for different activities that go beyond the level defined by the fluctuation-dissipation theorem. 
This has resulted into an unifying agent-based framework to model \emph{active matter}  \citep{Schweitzer2018}.
As other types of self-organizing systems, active matter \citep{bechinger2016active} relies on the take-up of energy that can be used for example for active motion or structure formation.
Our theoretical description is based on the coupling between \emph{driving} variables, $w_{i}(t)$, to describe the take-up, storage and conversion of energy, and \emph{driven} variables,  $x_{i}(t)$, to describe the energy consuming activities.
Modified Langevin equations reflect the stochastic dynamics of  both types of variables.  System-specific hypotheses about their coupling can be encoded in additional non-linear functions.

\subsection{Wisdom of crowds: Response to aggregated information}
\label{sec:wisdom-crowds}

The additive stochastic dynamics discussed above can be used to model the opinion dynamics of agents.
Here, $x_{i}(t)$ denotes the continuous opinion of agent $i$.
Let us consider the so-called \emph{wisdom of crowd} (WoC) effect, where the values of $x$ are usually mapped to the positive space, $x\geq 0$.
Agents are given a particular question with a definite answer, unknown to them, for example: ``What is the length of the border of Switzerland?'' \citep{Helbing2011}.
Their opinion $x_{i}>0$ then denotes their individual estimates about this length.
The WoC effect states that if one takes $N$ \emph{independent} estimates, the average $\mean{x}=(1/N)\sum_{i} x_{i}$ is close to the true value ${x}^{T}$.
I.e., the WoC effect is a purely statistical phenomenon, where the ``wisdom'' is on the population level.
It only works if the distribution of estimates $P(x)$ is sufficiently \emph{broad}, i.e. the variance, or the \emph{group diversity} of opinions, should be high.

In case of \emph{social influence}, e.g. information exchange between agents, estimates are no longer independent and the variance can reduce considerably.
One can argue that social influence could help agents to converge to a mean value closer to the truth.
On the other hand, social influence could also help agents to converge to a mean value much further \emph{away} from the truth -- without recognizing it.
That means, agents can collectively convince each other about the \emph{wrong} opinion to be the \emph{right} one.

Because this is a real-world problem for all social decision processes, it has been studied both experimentally \citep{Helbing2011,rauhut2011reply,Mavrodiev2013} and theoretically \citep{Mavrodiev2012}. 
In a controlled experiment, agents are given the same question a number of times.
They form an independent initial estimate $x_{i}(0)$.
After each subsequent time step, agents are given additional information about the estimates of other agents, which allows them to correct their own estimate, i.e. $x_{i}(t)$ becomes a function of time. 
This can be described by the dynamics:
\begin{equation}
  \label{eq:general}
  \frac{d x_{i}(t)}{dt}=  \gamma \left[ x_{i}(0)-x_{i}(t)\right] + \sum\nolimits_{j} \mathcal{F}(x_{j},x_{i}) + D \xi_{i}(t)
\end{equation}
which is a modification of Eqn. \eqref{eq:b}. 
The relaxation term $-\gamma x_{i}$ is now corrected by the initial estimate.
That means, without any social influence $x_{i}(t)$ has the tendency to converge to $x_{i}(0)$, rather than to zero.
$\gamma$ is the strength of an agent's \emph{individual conviction}.
Thus, the first term describes the tendency of an agent to stick to the original opinion.

$\mathcal{F}(x_{j},x_{i})$ describes the \emph{interaction} between agents, specifically between their opinions.
In a controlled experiment \citep{Helbing2011} agents can for example at each time step get information about the estimates $x_{j}(t)$ of \emph{all other agents} (full information regime), or only information about the \emph{average estimate} $\mean{x(t)}$ (aggregated information regime).
These regimes are expressed in different forms of $\mathcal{F}$.
A general ansatz reads:
\begin{align}
  \label{eq:12}
  \mathcal{F}(x_{j},x_{i})= \left[x_{j}(t)-x_{i}(t)\right] w_{ij}
\end{align}
Different forms of $w_{ij}$ encode how much weight agent $i$ attributes to the opinion of agent $j$. 
For the aggregated information regime, the quantity $w_{ij}$ is a constant, $w_{ij}={\alpha}/{N}$, where $\alpha$ denotes the strength of the \emph{social influence}.
With this, we can rewrite the dynamics of Eqn.~\eqref{eq:general} as:
\begin{align}
  \label{eq:agginformation}
  \frac{d x_{i}(t)}{dt} &=  \gamma \left[ x_{i}(0)-x_{i}(t)\right] + \alpha \left[\mean{x(t)}-x_{i}(t)\right]+ D \xi_{i}(t) \\
  &= - {\gamma}^{\prime} x_{i}(t) + \alpha\mean{x(t)} + \gamma x_{i}(0) + D \xi_{i}(t) \nonumber
\end{align}
with ${\gamma}^{\prime}=\gamma+\alpha$. 
Eqn. \eqref{eq:agginformation} highlights the \emph{coupling} to the \emph{``mean field''} formed by all agents,
because all agents interact and have the same $w_{ij}$.
The stochastic equation for the mean opinion \citep{Mavrodiev2012}: 
\begin{equation}
  \label{eq:mean}
  \frac{d \mean{x(t)}}{dt}=  \gamma \left[ \mean{x(0)}-\mean{x(t)}\right]
  + \frac{D}{\sqrt{N}} \mean{\xi(t)}
\end{equation}
describes a so-called \emph{Ornstein-Uhlenbeck process} and has an analytic solution. 

  \begin{figure}[htbp]
        \centering
\includegraphics[width=0.42\textwidth]{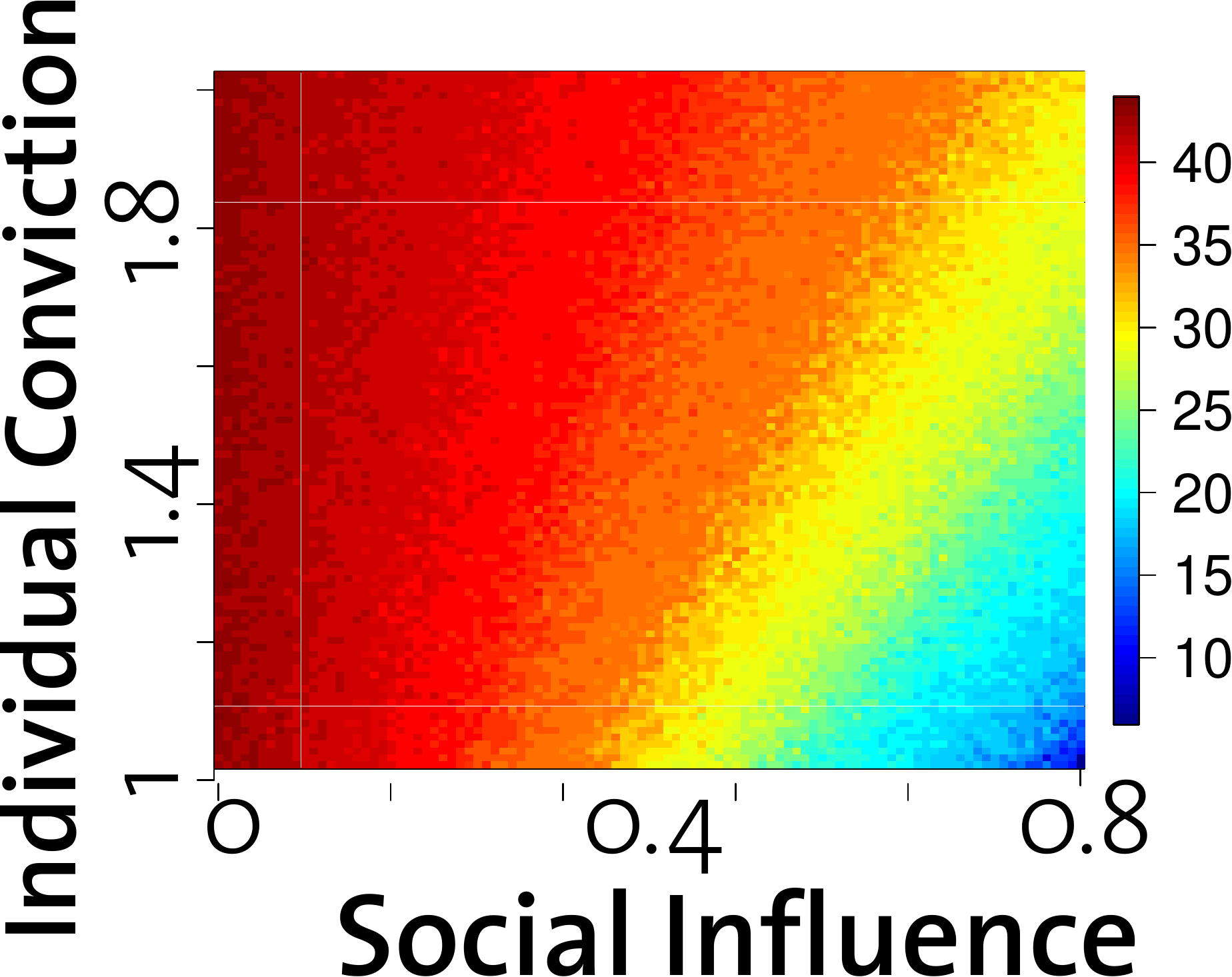}(a)
    \hfill
 \includegraphics[width=0.42\textwidth]{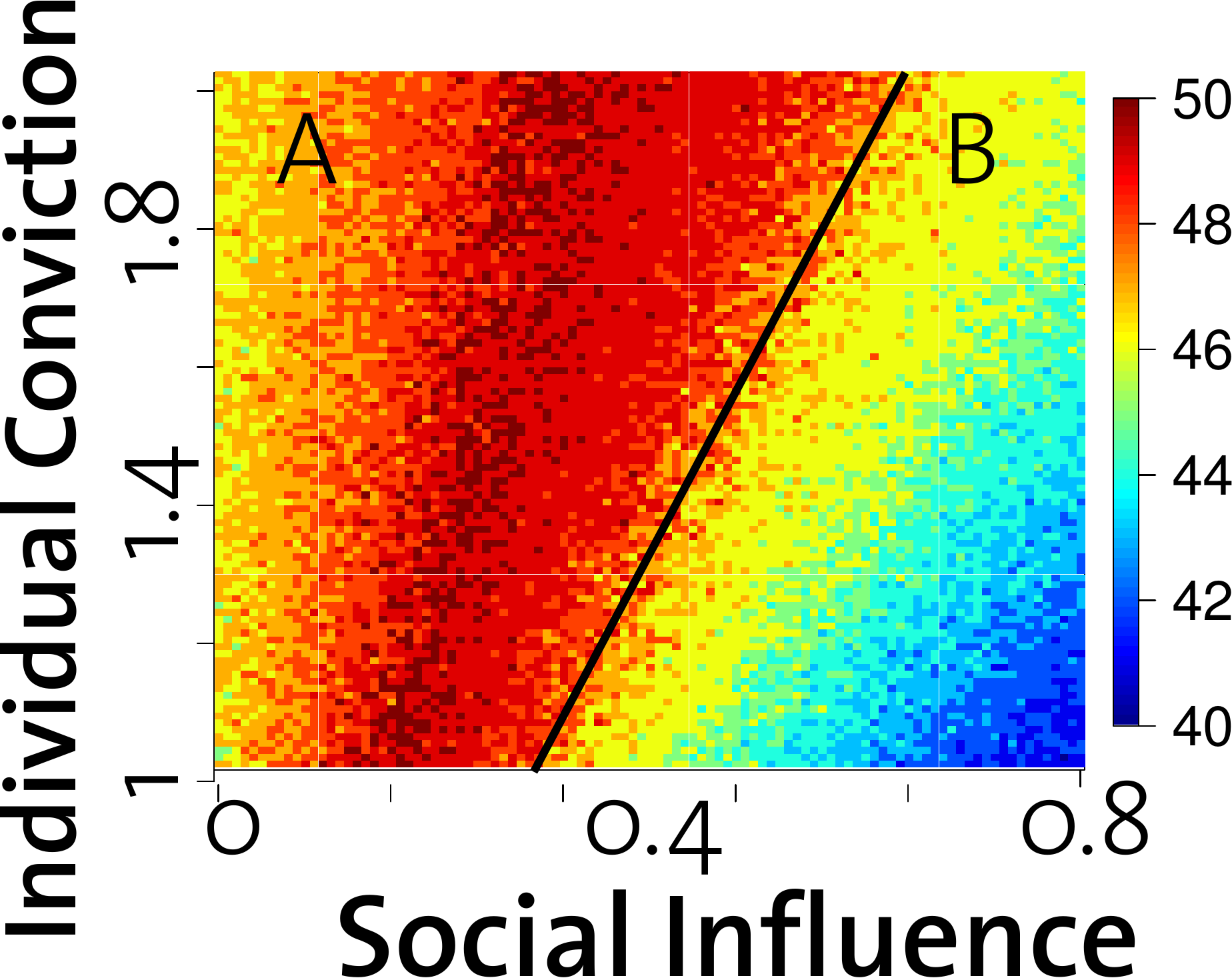}(b)
       \caption{WoC effect for a  parameter sweep of individual conviction $\gamma$ and social influence $\alpha$.
         Different initial conditions: (a) $\ln {x^{T}}< \mean{\ln x(0)}$, (b) $\ln {x^{T}}> \mean{\ln x(0)}$.
 The color code indicates how close the wisdom of crowds approaches the known truth: 50 (red) is the best performance.
\citep{Mavrodiev2012}}
\label{fig:woc}
\end{figure}
To better understand the role of the two model parameters, individual conviction $\gamma$ and social influence $\alpha$, one can run agent-based simulations with the dynamics of Eqn.~\eqref{eq:agginformation} \citep{Mavrodiev2012}.
Figure~\ref{fig:woc} shows how well the wisdom of crowds performs in reaching the (known) truth $x^{T}$, dependent on the \emph{initial conditions}, $x_{i}(0)$.
Figure~\ref{fig:woc}(a) illustrates a starting configuration, where $\ln {x^{T}}< \mean{\ln x(0)}$.
In this case, a \emph{larger} social influence \emph{always} leads to a \emph{worse} performance of the WoC (indicated by the monotonous color change from red to blue).
Figure~\ref{fig:woc}(b), on the other hand, illustrates a starting configuration, where $\ln {x^{T}}>\mean{\ln x(0)}$.
Here we find instead a \emph{non-monotonous} color change.
While for very small values of the social influence the performance is lower (yellow), it increases for medium values of $\alpha$ (red), before it declines again for large values of $\alpha$ (blue).
Hence, in a region $A$ increasing social influence can help the wisdom of crowds, while in a region $B$  increasing social influence rather distorts the WoC effect.

\paragraph{Applications. }

It is quite remarkable how well the agent-based dynamics of Eqn. \eqref{eq:agginformation} describes the empirical results for the aggregated information regime \citep{Mavrodiev2013} obtained in controlled experiments with humans \citep{Helbing2011}. 
It was found that the adjustment of individual opinions depends \emph{linearly} on the distance to the \emph{mean} of all estimates, even though the correct answers for different questions differ by several orders of magnitude.

\subsection{Bounded confidence: Consensus versus coexistence of opinions}
\label{sec:bounded-confidence}

The redistribution model and the WoC model both assume that agents in a population interact via an aggregated variable. 
This is different in the so-called ``bounded confidence'' model \citep{Lorenz2007}. 
Here, the continuous values of $x$ represent opinions which are mapped to the positive space, $x\geq 0$, and transformed to the unit interval $[0,1]$.
The model assumes that two randomly chosen agents $i$ and $j$ can interact  only if they are sufficiently close in their
values $x_{i}(t)$, precisely if $\abs{x_{i}(t)-x_{j}(t)}<\epsilon$,
i.e. below a given threshold $\epsilon$, which defines a \emph{tolerance} for other's opinions.
Thanks for their interaction, agents adjust their opinions
towards each other, which  is motivated by social arguments:
\begin{equation}
  \label{eq:mean}
  \frac{d x_{i}(t)}{dt}= \gamma \left[x_{j}(t) - x_{i}(t)\right] \Theta[z_{ij}(t)]\;;\quad
  z_{ij}(t)= \epsilon - \abs{x_{j}(t)-x_{i}(t)}
\end{equation}
Here, $\Theta[z]$ is the Heaviside function, which returns $\Theta[z]=1$ if $z\geq 0$ and $\Theta[z]=0$ otherwise.
The parameter $0<\gamma\leq 0.5$ basically defines the time scale at which the opinions of the two agents converge, provided that $z_{ij}(t)\geq 0$.
If $\gamma=0.5$, both agents immediately adjust their $x_{i}(t)$, $x_{j}(t)$ towards the common mean.

\begin{figure}[htpb]
  \centering
  \includegraphics[width=0.45\textwidth]{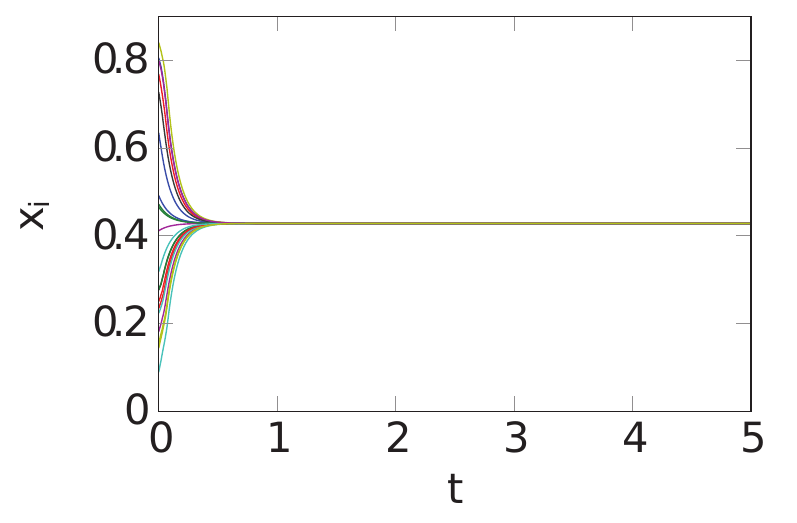}(a) \hfill
\includegraphics[width=0.45\textwidth]{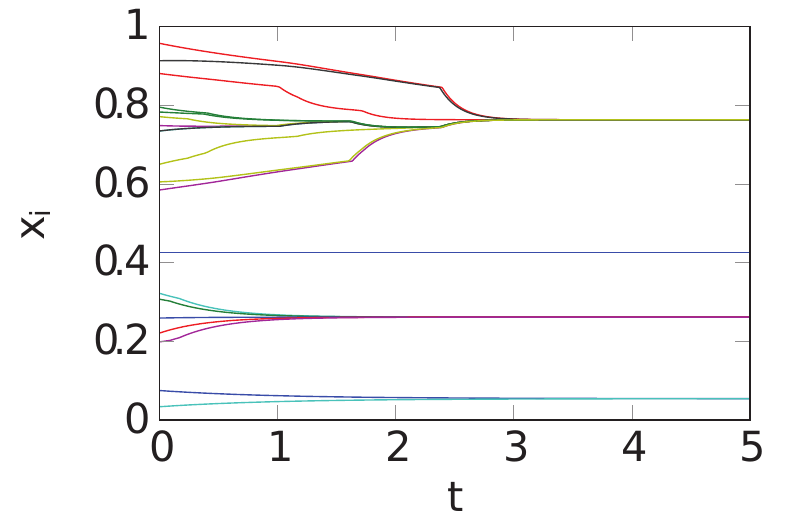}(b)
  \caption{Bounded confidence dynamics, Eqn.~\eqref{eq:mean} for $\gamma$=0.5 and different threshold values: (a) $\epsilon$=0.5, (b) $\epsilon$=0.1}
  \label{fig:bond}
\end{figure}
Because at each time step only two randomly chosen agents can interact, the sequence of interactions matters for the final outcome and the \emph{collective opinion dynamics} becomes a path dependent process. 
The main research question addressed with this type of model is about \emph{consensus formation}, i.e. about the conditions under which a population of agents with randomly chosen initial opinions $x_{i}(0)\in[0,1]$ converges to \emph{one} final opinion.
While $\gamma$ only determines the time scale for convergence, the threshold $\epsilon$ mainly decides about the outcome. 
Fig. \ref{fig:bond} shows examples for  two different values of $\epsilon$.
For $\epsilon=0.5$, we indeed find consensus, while for $\epsilon=0.1$ we observe the \emph{coexistence of two final opinons}.
It was shown \citep{deffuant2000mixing} that convergence towards consensus can be expected for $\epsilon\geq
0.25$. If the interaction threshold is below this critical value, we observe instead
the convergence towards multiple stationary opinions.
This reminds on
the period doubling scenario, i.e. the multiplicity of solutions found for the logistic map, Eqn. \eqref{eq:3}, when varying the
control parameter $r$.
Similar to that example, in the bounded confidence
model we also observe ``windows'' in which convergence to one stationary
value is observed.

Instead of a sequence of dyadic interactions of agents, one can also consider \emph{group interactions}, in which many agents interact simultaneously.
The dynamics then changes into: 
\begin{equation}
  \label{eq:mean-g}
  \frac{d x_{i}(t)}{dt}=\frac{\gamma}{\mathcal{N}_{i}(\epsilon,t)} \sum\nolimits_{j} \left[x_{j}(t) - x_{i}(t)\right] \Theta[z_{ij}(t)]\;;\quad
 \mathcal{N}_{i}(\epsilon,t)= \sum\nolimits_{j} \Theta[z_{ij}(t)]
\end{equation}
where the normalization $\mathcal{N}_{i}$ depends on all agent's opinions and therefore on time, but also on the threshold $\epsilon$.
For $\epsilon \to 1$, $\mathcal{N}_{i}(\epsilon,t) \to N$, and we obtain again an agent dynamics which is coupled to the \emph{mean}, with $\gamma^{\prime}={\gamma}/{N}$:
\begin{equation}
  \label{eq:mean2}
  \frac{d x_{i}(t)}{dt}= -\gamma^{\prime} x_{i}(t) + \gamma^{\prime} \mean{x(t)}
\end{equation}

\paragraph{Applications. }

\begin{figure}[htbp]
  \centering
\includegraphics[width=0.45\textwidth]{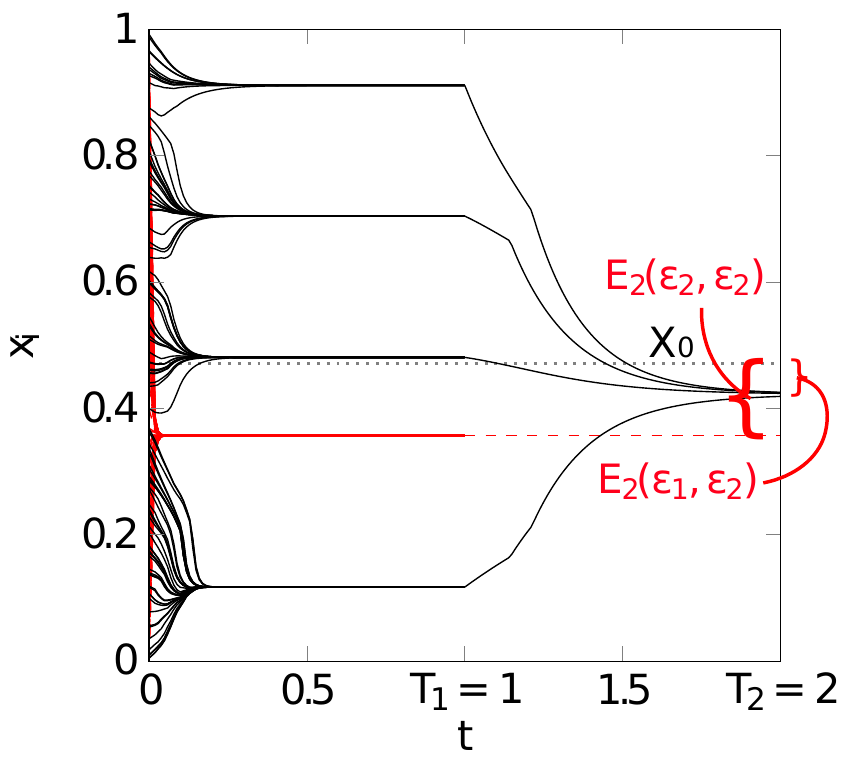}(a)\hfill
  \includegraphics[width=0.40\textwidth]{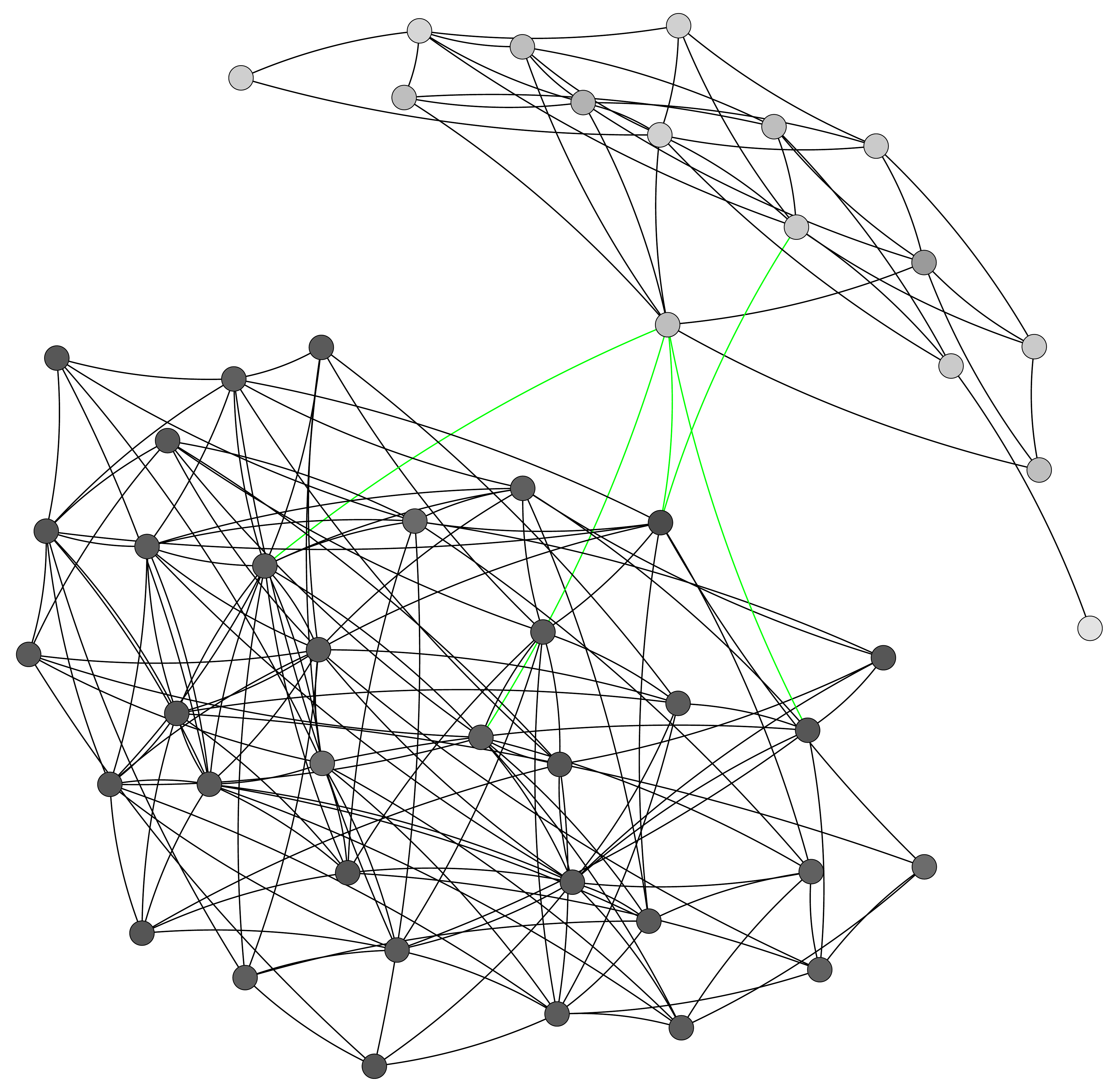}(b)
  \caption{(a) Hierarchical opinion dynamics with $\epsilon_{1}=0.1$, $\epsilon_{2}=1$. Additionally  in the dynamics an asymmetric preference for opinions closer to zero is assumed  \citep{Pfitzner2013}. (b) Opinion dynamics with in-group influence, Eqs.~\eqref{eq:13}, \eqref{eq:14}, with $\epsilon$=0.3. Green links indicate that agents would \emph{not} interact \emph{without} the influence of their in-groups \citep{Groeber2009}.}
  \label{fig:bounded2}
\end{figure}

Most applications of the bounded confidence model propose ways to \emph{enhance consensus formation}, for instance by introducing \emph{asymmetric} confidence values $\epsilon^{\mathrm{left}}$, $\epsilon^{\mathrm{right}}$ \citep{hegselmann2002opinion}.
Consensus can be also fostered using a \emph{hierarchical} opinion dynamics \citep{Pfitzner2013,Pfitzner2012} as shown in Figure~\ref{fig:bounded2}(a).
During a first time period, all agents adjust their opinions according to the bounded confidence model, Eqn.~\eqref{eq:mean}, such that groups with distinct opinions are formed.
During a second period, these group opinions are represented by \emph{delegates} that follow the same dynamics, but have a larger threshold than ``normal'' agents, $\epsilon_{2}>\epsilon_{1}$.
Therefore, these delegates are likely to find a consensus even in cases where the original agent population fails to converge to a joint opinion.

Another application of the bounded confidence model explains the emergence of so-called \emph{local cultures}, a commonly shared behavior within a cluster of firms \citep{Groeber2009}.
The basic assumption is that agents \emph{keep partnership relations} from past interactions and this way form so-called \emph{in-groups} $I_{i}(t)$.
The opinions of agents from the in-group continue to influence an agent's opinion, this way leading to an \emph{effective opinion}
\begin{align}
x_i^{\textrm{eff}}(t) =& \left[1-\alpha_i(t)\right]x_i(t) + \alpha_i(t) \mean{x}_i^I(t) 
                   \label{eq:13}
\end{align}
Here $\mean{x}_i^I(t)$ is the mean opinion of agents in the in-group of $i$, and $\alpha_{i}(t)$ weights this influence against the ``native'' opinion $x_{i}(t)$ of agent $i$, considering the size of the in-group, $\abs{I_{i}(t)}$:
\begin{align}
  \label{eq:14}
  \mean{x}_i^I(t) = & \frac{1}{\abs{I_{i}(t)}}\sum_{j\in I_{i}(t)} x_{j}(t)		\;;\quad \alpha_i(t) = \frac{|I_i(t)|}{|I_i(t)|+1}
\end{align}
While agents adjust their opinions $x_{i}(t)$ according to the bounded confidence model, Eqn. \eqref{eq:mean}, their effective opinions $x_{i}^{\mathrm{eff}}(t)$ decide about their interactions, i.e.
$z_{ij}(t)= \epsilon - \abs{x^{\mathrm{eff}}_{j}(t)-x^{\mathrm{eff}}_{i}(t)} $
Only if interaction takes place, i.e. $z_{ij}(t)\geq 0$, $j$ is added to the in-group of $i$ and a link between agents $i$ and $j$ is formed.
Because a change of $\mean{x}_{i}^{I}(t)$ can occur even if $i$ does not interact, this impacts $x_{i}^{\mathrm{eff}}(t)$ continuously.
So, two agents $i$ and $j$ randomly chosen at different times may form a link later, or may remove an existing link because of their  
in-groups' influence, as illustrated in Figure~\ref{fig:bounded2}(b).
This feedback between agents' opinions  and their in-group structure sometimes allows to obtain consensus, or a common ``local culture'', even in cases where the original dynamics would fail.

\subsection{Bilateral encounters:  Reputation growth from battling}
\label{sec:battling}

In opinion dynamics, the variable $x_{i}(t)$ does not assume any intrinsic value, i.e. it is not favorable to have a larger or smaller $x_{i}(t)$.
This changes if we consider that $x_{i}(t)$ represents the \emph{reputation} of agent $i$, where ``higher'' means ``better''.
Reputation, in loose terms, summarizes the ``status'' of an agent, as perceived by others.
It can be seen as a social capital and influences for example the choice of interaction partners.
For firms, reputation is an \emph{intangible asset}, that means, it is difficult to quantify, but at the same time influences the decisions of investors or customers \citep{Schweitzer2019}.
Even if the measurement of reputation is a problem, it is obvious that reputation has to be maintained, otherwise it fades out.
This can be captured by the dynamics already discussed:
\begin{equation}
  \label{eq:15}
  \frac{d x_{i}(t)}{dt}=  - \gamma x_{i}(t) + \sum\nolimits_{j} \mathcal{F}(x_{j},x_{i}) 
\end{equation}
The multiplicative term describes the exponential decay of reputation over time.
To compensate for this  requires a continuous effort, expressed in the interaction term $\mathcal{F}(x_{j},x_{i})$.
This assumes that reputation can be (only) built up in interactions with other agents $j$.
One could include into the dynamics of Eqn. \eqref{eq:15} another source term for reputation which solely depends on the efforts of agent $i$, but its justification remains problematic.
One could argue that, for example, the reputation of scientists depends on their effort writing publications.
But this individual effort can hardly be quantified and compared across scientists.
More importantly, not the effort matters for the reputation, but the attention the publication receives from \emph{other} scientists \citep{Pfitzner2014}, as quantified e.g.
by the number of citations. 
Eqn. \eqref{eq:15} is similar to the general dynamics proposed in Eqn.~\eqref{eq:general}, just that no additive stochastic influence is explicitly considered here and no intrinsic reputation $x_{i}(0)$ is assumed.
For the interaction term, we can in the following separately discuss two different \emph{limit cases}:
(i) Reputation is obtained solely during direct battles between two agents, where the winner gains in reputation and the looser not. 
(ii) Reputation is obtained solely from interacting with other agents and increases with their reputation.

In case of individual battles, we assume that during each time step each agent $i$ has a \emph{bilateral} interaction with any other agent $j$.
For the interaction term we propose \citep{Perony2019}: 
\begin{align}
  \label{eq:44}
\mathcal{F}(x_{j},x_{i}) = \frac{1}{N}\sum\nolimits_{j}\rho(x_{i},x_{j})  \Big\{g+ h \Delta_{ji}\, \Theta\left[\Delta_{ji} \right]\Big\} \;; \quad \Delta_{ji}={x_{j}(t)}-{x_{i}(t)}  
\end{align}
$\rho(x_{i},x_{j})$ is a function that decides which agent will be the winner in an interaction between any two agents  $i$ and $j$.
It depends on the reputation of both agents, but additionally also considers random influences \citep{Perony2019}.
The expression in curly brackets determines the reputation gain for the winner.
It consists of two contributions.
$g$ is a constant reward for every winner. It accounts for the fact that engaging in such fights is a costly action that should be compensated. 
$h\Delta_{ji}$ is a bonus reward that applies only if agent $i$ was the one with the lower reputation, $x_{i}<x_{j}$, and still won the fight.
This is expressed by the Heavyside function $\Theta[\Delta_{ji}]$.
Note that, in this model, only the reputation of the \emph{winning agent} will be changed, the loosing agent does not additionally loose in reputation.

\begin{figure}[htbp]
  \centering
  \includegraphics[width=0.45\textwidth]{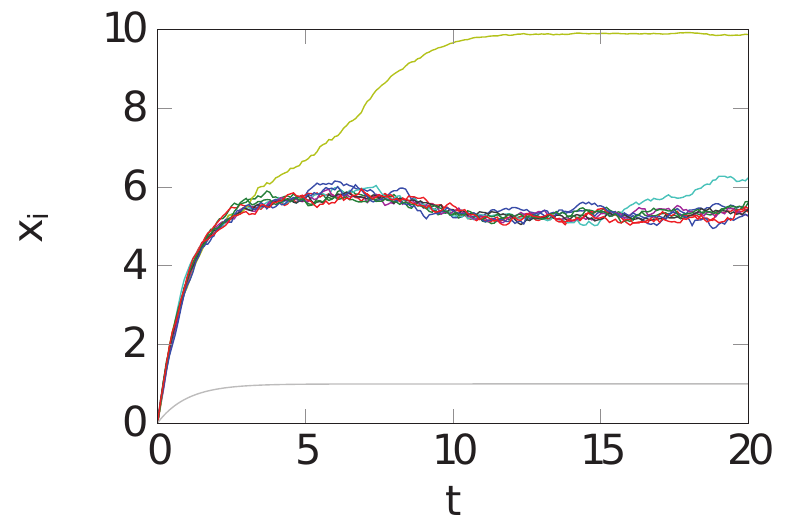}(a)\hfill
    \includegraphics[width=0.45\textwidth]{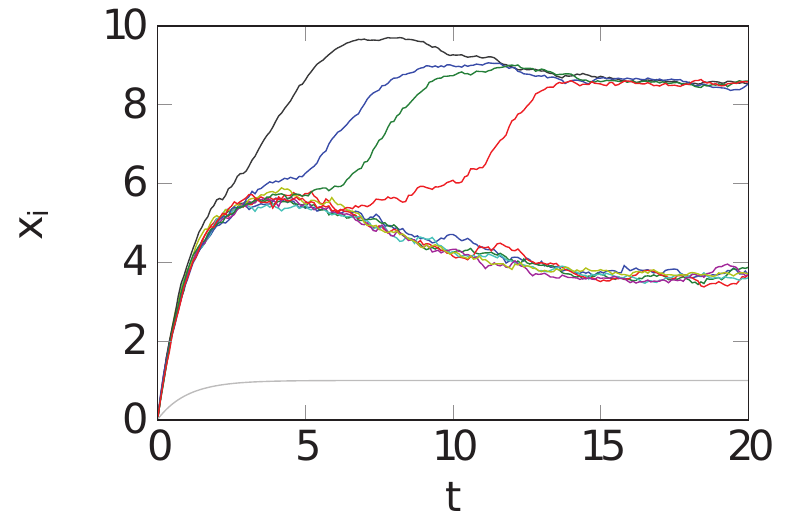}(b)
    \caption{Reputation dynamics from bilateral encounters, Eqs.~\eqref{eq:15}, \eqref{eq:44}. Variance of the normal distribution, from which random influences are drawn: (a) $\sigma^{2}$=1.4, (b) $\sigma^{2}$=1.2 \citep{Perony2019}. }
    \label{fig:battle}
\end{figure}

\paragraph{Applications. }

Fights between individuals are ubiquitous in the animal kingdom to establish reputation.
In a biological setting, reputation differences translate into  \emph{dominance relations}.
Hence, this model has a particular relevance to explain social hierarchies in animal societies \citep{Bonabeau1999}.
It allows to test  whether hierarchies in social organizations are an \emph{emerging phenomenon} or whether they result from the reinforcement of \emph{intrinsic advantages} of individuals.
Subsequently, an interaction model allows to test different feedback mechanisms.
If the winner is rewarded \emph{and} the looser is punished, this results in a double reinforcement, and the model displays a strong \emph{lock-in effect}.
That is, the outcome is almost entirely determined by the first few interactions, initial random differences are just amplified.
To obtain realistic hierarchies, it is sufficient to only reward the winner  \citep{Perony2019}. 
For hierarchies with different levels the mentioned function $\rho(x_{i},x_{j})$ plays a particular role.
It reflects random  influences, the magnitude of which is expressed by the variance $\sigma^{2}$ of a normal distribution.  
For large values of $\sigma^{2}$ egalitarian regimes are obtained, for intermediate values one agent dominates (despotic hierarchy), as shown in Figure~\ref{fig:battle}(a),
 while for small values layered hierarchies can be obtained as shown in Figure~\ref{fig:battle}(b)

\subsection{Network interactions:  Reputation growth through feedback cycles}
\label{sec:netw-inter-reput}

The second limit case, where reputation is obtained solely from interacting with other agents, can be expressed by the following interaction term:
\begin{equation}
\mathcal{F}(x_{j},x_{i}) =\sum\nolimits_{j} l_{ji}\, x_j(t) 
  \label{eq:dyn}
\end{equation}    
The coefficients $l_{ji}$ are unweighted, but directed \emph{links} between agents $j$ and $i$.
$l_{ji}(t)=1$ if there is a link from $j$ to $i$, i.e. agent $j$ can boost the reputation of $i$ proportional to its own reputation.
This is a very common feedback mechanism, also used to define \emph{eigenvector centrality} \citep{bonacich1987power}, with many applications.  
For example, in an online social network (OSN) like \texttt{Twitter}, a link $j\to i$ indicates that $j$ is a \emph{follower} of $i$, and the prominence of $j$ impacts the prominence of $i$.
$l_{ji}(t)=0$ if no directed link exists, and $l_{ii}(t) = 0$ because an agent cannot boost its own reputation.

With these considerations, the multi-agent system can be represented as a \emph{complex network}, $G(E,V)$ (graph), where \emph{nodes}, $V$ (vertices), represent agents and directed \emph{links},  $E$ (edges), between nodes their directed interactions.
The network structure is then encoded in an \emph{adjacency matrix} $\mathbf{A}$ with matrix entries $l_{ji}$.
Using the interaction term, Eqn.~\eqref{eq:dyn}, the \emph{stationary solution} for
the dynamics of Eqn.~\eqref{eq:15} can be formally written as:
\begin{equation}
  \label{eq:16}
  x_{i}^{\mathrm{stat}}= \frac{1}{\gamma}\sum\nolimits_{j} l_{ji} x_j^{\mathrm{stat}}
\end{equation}
This defines a set of coupled equations and has the structure of an eigenvalue problem.
It has a stationary solution only if the factor $\gamma$ is the \emph{eigenvalue} of the adjacency matrix $\mathbf{A}$.
That means, for arbitrarily chosen values of $\gamma$ different from an eigenvalue, the $x_{i}(t)$ will either grow too fast  (small $\gamma$) or too slow (large $\gamma$) to be balanced by the other $x_{j}(t)$, this way resulting in a non-stationary solution.
For a stationary solution, usually the \emph{largest} eigenvalue is taken because it guarantees that all solutions are positive (if the matrix $\mathbf{A}$ is non-negative).

One can eliminate $\gamma$ by transforming the absolute reputation values $x_{i}(t)$ into  \emph{relative reputations},  $y_{i}(t)=x_{i}(t)/\sum_{j}x_{j}(t)$.
This also has a practical implication: absolute values are hard to know and, to \emph{compare} agents, relative values are sufficient. 
Under most practical circumstances, however, one would also not be able to obtain a complete normalization, $\sum_{j}x_{j}(t)$.
But it is sufficient \citep{mavrodiev2019} if we can normalize by the largest reputation: $y_{i}(t)=x_{i}(t)/x_{z}^{\mathrm{max}}(t)$. 
\begin{equation}
\label{eq:rel-reputation}
\dfrac{d y_{i}}{dt} = \sum\nolimits_{j}l_{ji}\, y_{j}(t) - y_{i}(t)\sum\nolimits_{j} l_{jz}\, y_{j}(t)
\end{equation}
where $z$ is the index of the agent with highest absolute reputation
$x^{\mathrm{max}}_{z}(t)$ at time $t$, which is for instance often known in a OSN.
Its scaling impact on the relative reputation is summarized in the second term of Eqn.~\eqref{eq:rel-reputation}.
This represents the reputation \emph{decay} for agent $i$ with a strength equal to the total boost in reputation
that agent $z$ receives.
One can show \citep{mavrodiev2019} that an equilibrium solution for $x_{i}(t)$, Eqs.~\eqref{eq:15}, \eqref{eq:dyn}, is
also an equilibrium solution for $y_{i}(t)$ (with either normalization) up to a scaling factor.
Specifically, for an eigenvector $\mathbf{y}^{\lambda}$ of the adjacency matrix $\mathbf{A}$, the corresponding eigenvalue $\lambda$ is given by: $\sum_{j}l_{jz}\,y_{i}=\lambda$.

\begin{figure}[htbp]
  
  \begin{minipage}{.3\textwidth}
(a)    \includegraphics[height=3cm]{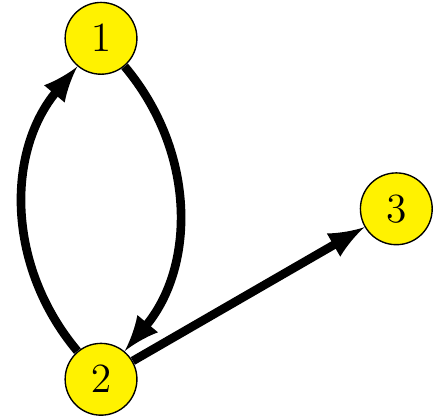}
    \begin{displaymath}
      A=\left(
        \begin{array}{ccc}
          0 & 1 & 0 \\
          1 & 0 & 1 \\
          0 & 0 & 0 \\
        \end{array}
      \right)
    \end{displaymath}

    \centerline{\includegraphics[width=.99\textwidth,angle=0]{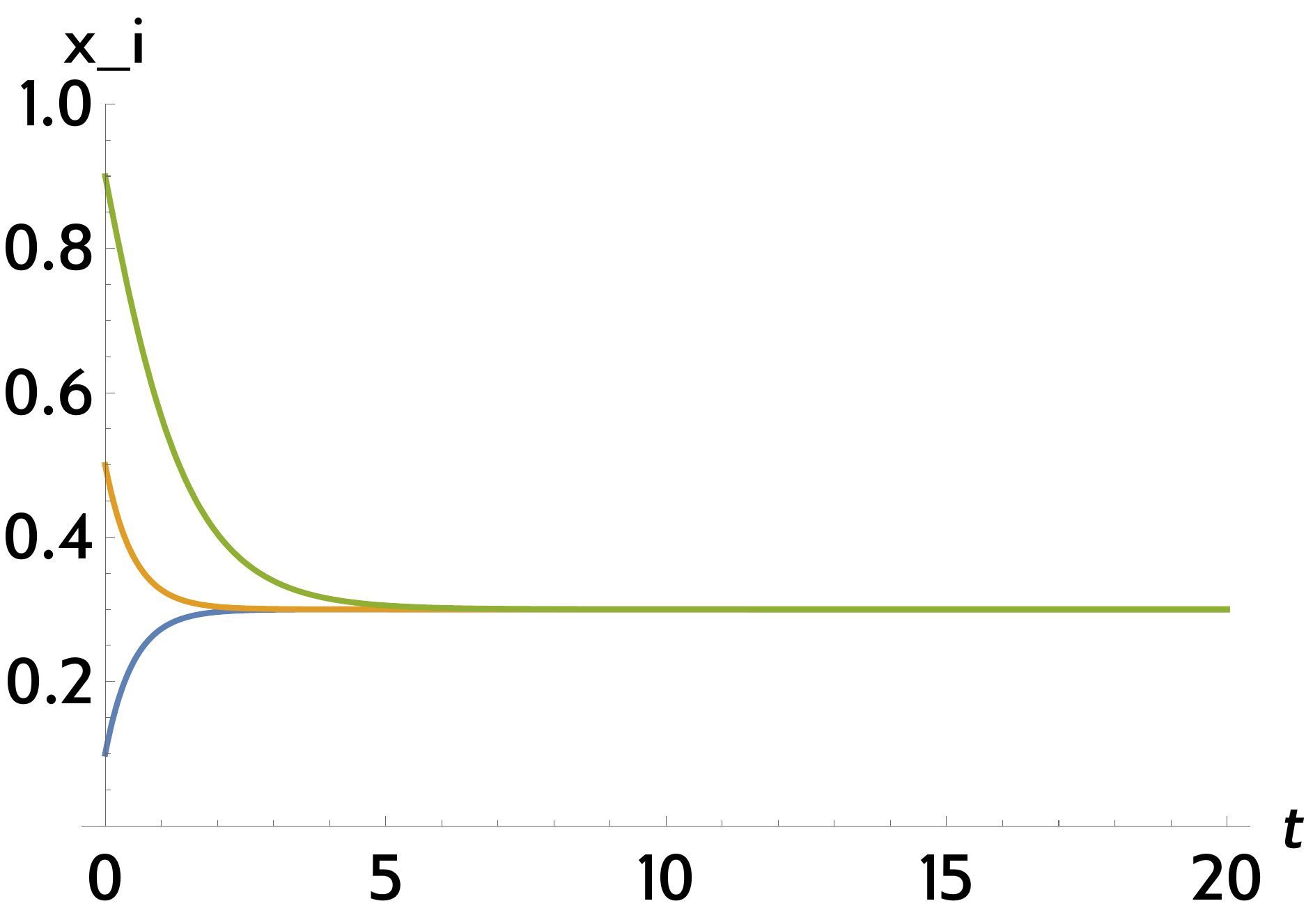}} \end{minipage}\hfill\vline\hfill
\begin{minipage}{.3\textwidth}
(b)$\quad$ \includegraphics[height=3cm]{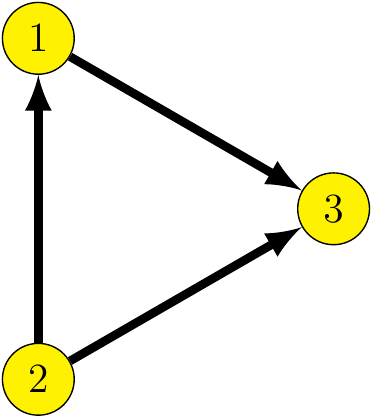}
\begin{displaymath}
A=\left(
\begin{array}{ccc}
0 & 0 & 1 \\
1 & 0 & 1 \\
0 & 0 & 0 \\
\end{array}
\right)
\end{displaymath}

    \centerline{\includegraphics[width=0.99\textwidth,angle=0]
        {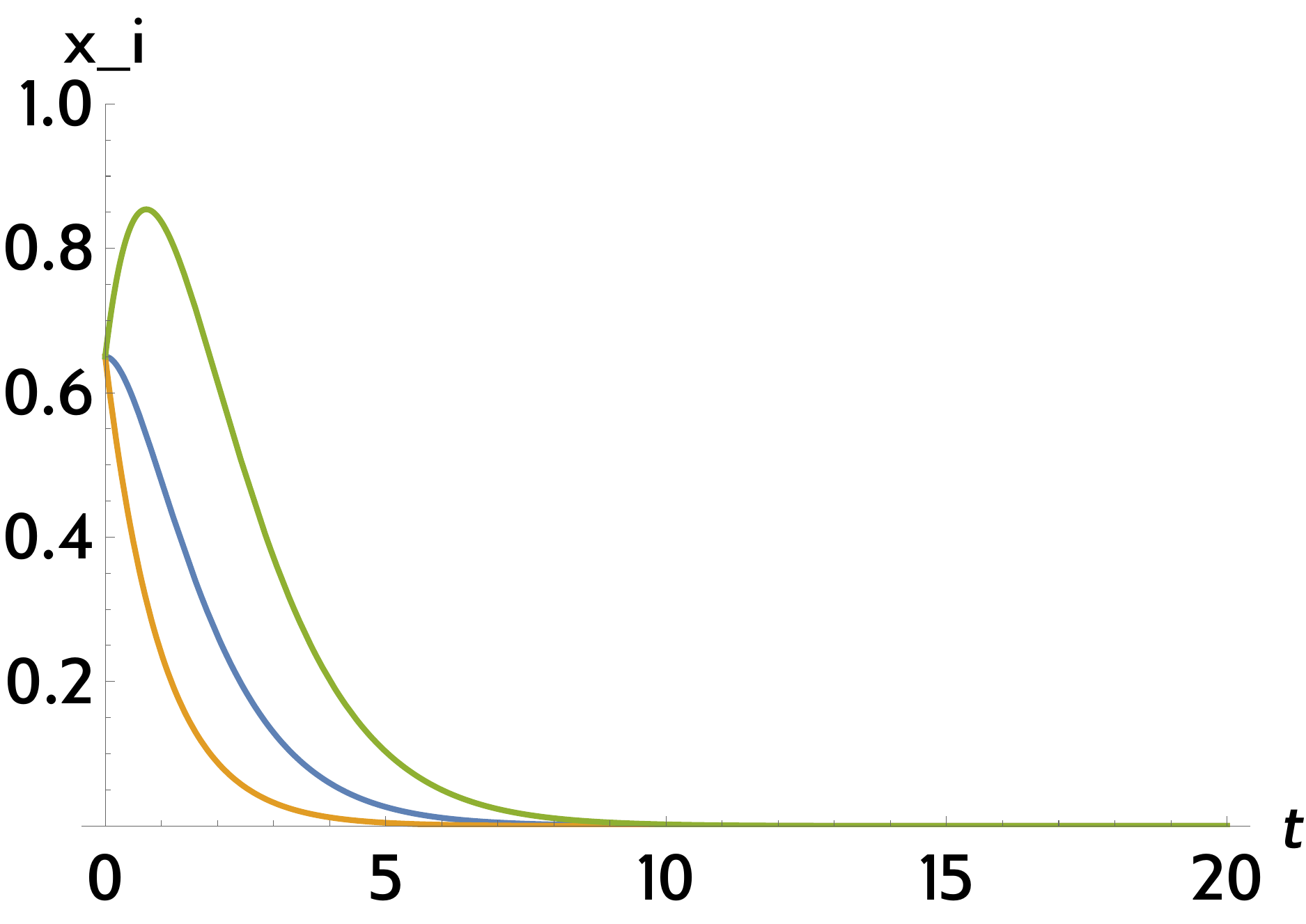}} \end{minipage}\hfill\vline\hfill
\begin{minipage}{.3\textwidth}
 (c) \includegraphics[height=3cm]{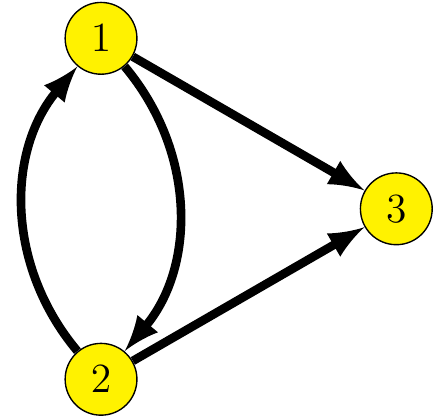}
\begin{displaymath}
A=\left(
\begin{array}{ccc}
0 & 1 & 1 \\
1 & 0 & 1 \\
0 & 0 & 0 \\
\end{array}
\right)
\end{displaymath}

    \centerline{\includegraphics[width=0.99\textwidth,angle=0]
        {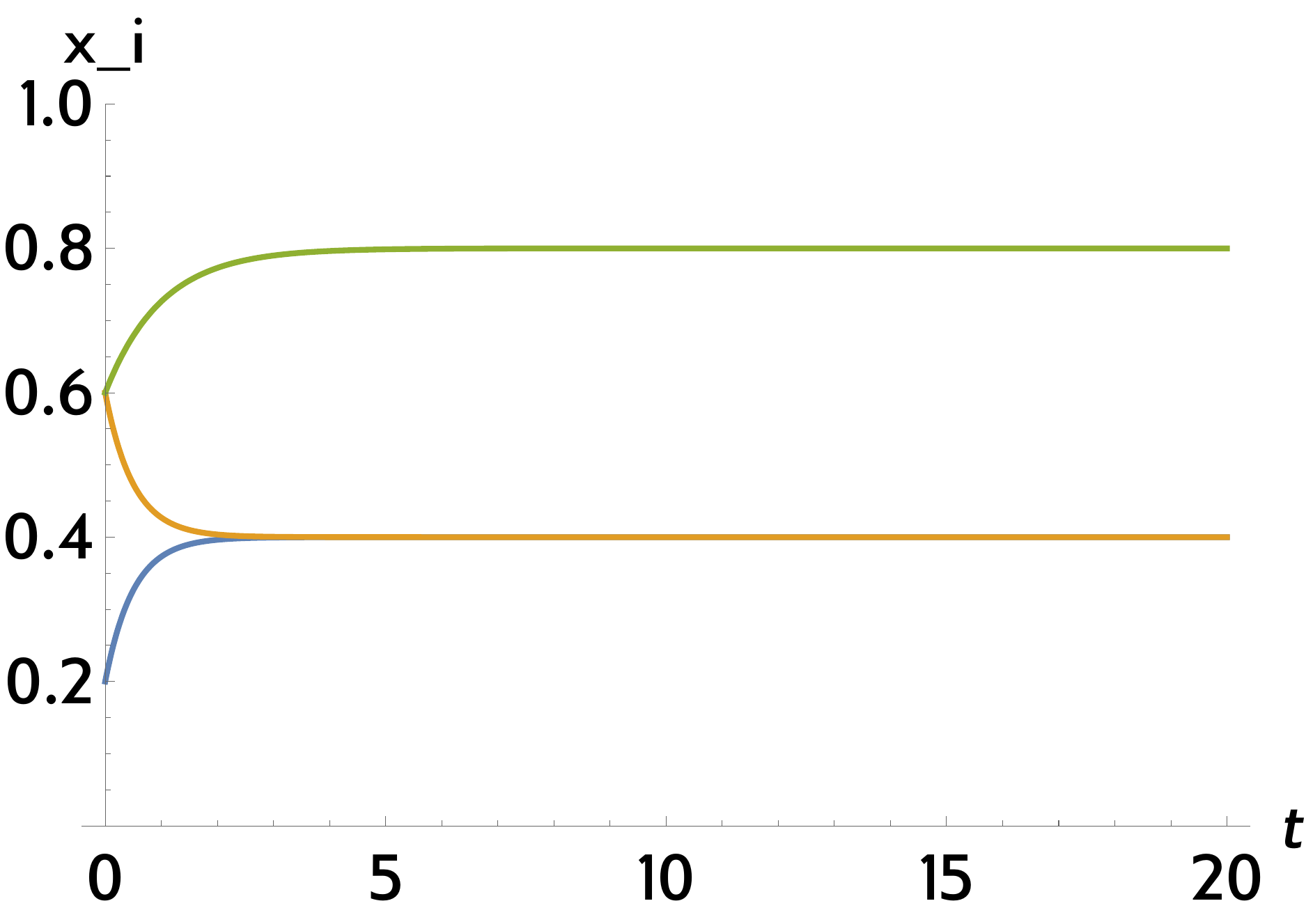}} \end{minipage}
\caption{Impact of the adjacency matrix on the reputation $x_{i}(t)$ of three agents.
  Only if cycles exist and agents are connected to these cycles, a non-trivial stationary reputation can be obtained.
  The chosen $\gamma=1$ is indeed an eigenvalue of the adjacency matrix in  (a) and (c). This is not the case for (b), hence we do not observe a stationary solution, just a convergence to zero. 
}
\label{fig:3examples}
\end{figure}

Whether \emph{non-trivial solutions} for the stationary reputation values, $x_{i}(t)\to x_{i}^{stat}$, exist, strongly depends on the \emph{adjacency matrix}, as illustrated in Figure~\ref{fig:3examples}.
Specifically, if an agent has no incoming links that boosts its reputation, $x_{i}(t)$ will go to zero.
Therefore, even if this agent has an outgoing link to other agents $j$, it cannot boost their reputation.
Non-trivial solutions depend on the existence of \emph{cycles} which are formally defined as subgraphs in which there is a closed path from every node in the subgraph back to itself.
The shortest possible cycle involves two agents, $1\to 2 \to 1$.
This maps to \emph{direct reciprocity}: agent 1 boosts the reputation of agent 2, and vice versa.
Cycles of length 3 map to \emph{indirect reciprocity}, for example $1\to 2 \to 3 \to 1$.
In this case, there is \emph{no} direct reciprocity between any two agents, but all of them benefit regarding their reputation because they are part of the cycle.
In order to obtain a non-trivial reputation, an agent not necessarily has to be part of a cycle, but it has to be connected to a cycle. 

\paragraph{Applications. }

The type of feedback dynamics discussed above is widely used to characterize the importance of nodes in a network.
Already \texttt{Google's} early version of the \texttt{PageRank} algorithm built on this.
A related measure, \emph{DebtRank} was introduced to quantify the importance of institutions in a financial network \citep{Puliga2012}. 
Further, the approach has found an important application to model online social networks (OSN) \citep{mavrodiev2019}. 
For example in \texttt{Twitter} or \texttt{Instagram}, the reputation of users is not just determined by the \emph{number} of followers, but also by their \emph{reputation}. 
It makes all the difference, whether individual $i$ is a follower of the famous actor $z$, or the other way round.

Social networks are often characterized by a \emph{core-periphery} structure, where the core contains a subset of well connected users.
This is important for the application of this model, as it relies on the existence of cycles.
These cycles can be of any length, even structures of interlocking cycles can be present.
Their existence, as reflected in the \emph{adjacency matrix}, then impacts the corresponding eigenvalues and hence the (relative) reputation of users. 
It is computationally hard to detect such interlocking cycle structures in real social networks.
In a case study of 40 million \texttt{Twitter} users, reputation was therefore measures by means of a \emph{D-core decomposition} \citep{Mavrodiev2016}.

\section{Growth combined with network dynamics}
\label{sec:feedback}

\subsection{Nonlinear growth of knowledge stock:  Entry and exit dynamics}
\label{sec:nonlinear-growth}

Not only reputation depends on the feedback from other agents, also knowledge growth crucially relies on it.
Let us assume that the quantity $x_{i}(t)$ now describes the \emph{knowledge stock} of agent $i$, for example the R\&D (research and development) experience of a firm, measurable by its number of patents and research alliances.

The value of knowledge continuously decreases if it is not maintained.
Hence, we can propose the same general Eqn.~\eqref{eq:15} for reputation to also describe the dynamics of the knowledge stock.
To compensate for the decay, we assume that the growth of knowledge is mainly driven by input from \emph{other agents}, i.e. by \emph{R\&D collaborations}, rather than by own activities.
This reflects empirical observations for \emph{innovation networks} of firms \citep{Tomasello2016}.

Different from reputation growth, for which no upper limit needs to be assumed, it is reasonable to consider a \emph{saturation} for the growth of knowledge stock, similar to the quadratic term used in the saturated growth Eqn.~\eqref{eq:xsat}.
At higher levels of knowledge stock, it becomes more difficult to ``absorb'' new knowledge, i.e. to incorporate it into a firm, simply because of the internal complexity associated with the way knowledge is stored and linked internally.
Because of this absorptive capacity, and we propose the following dynamics for the knowledge stock \citep{Battiston2009}:
\begin{equation}
  \frac{dx_{i}(t)}{dt}= - \gamma\, x_{i}(t) + \nu \sum\nolimits_{j} l_{ji}\, x_j(t) + \nu^{\mathrm{ext}} \sum\nolimits_{j} p_{ji}\, x_j(t) - \kappa \sum\nolimits_{j} l_{ij}\, x_{i}^2(t)
  \label{eq:18}
\end{equation}
The knowledge growth is mainly determined by the knowledge stock of agents $j$ that have \emph{direct link} to agent $i$, as expressed by the $l_{ji}$.
But we can additionally consider that some links, denoted by $p_{ji}$, provide direct input to $i$ from particular valuable agents.
For example, instead of obtaining \emph{indirect} knowledge input from an agent $k$ via other agents $j$, agent $i$ would much more benefit if $k$ had a \emph{direct} link to $i$.
So, if $p_{ji}=1$, in addition to the usual benefit $\nu$ there will be an extra benefit  $\nu^{\mathrm{ext}}$ from interacting with this valuable agent.
As we have already noticed the importance of \emph{indirect reciprocity} in the growth of $x_{i}$, such extra benefit could also arise from links that contribute to closing \emph{cycles} in the interaction network.
This would allow feedback cycles for instance in the development of a technology. 

With or without the additional saturation and growth terms, under certain conditions for the parameters Eqn.~\eqref{eq:18} will lead to a \emph{stationary} solution for the knowledge stock of all agents, as discussed in Sect. \ref{sec:netw-inter-reput}.
An evaluation of the stationary solutions of Eqn.~\eqref{eq:18} that corresponds to Figure~\ref{fig:3examples} can be found in \citep{Battiston2009}. 
To make the dynamics of the system more realistic, we can further consider an \emph{entry and exit} dynamics.
Not successful agents may leave the system, whereas new agents enter.
This is associated with \emph{rewiring the network} that represents the collaboration interactions, i.e. some links are removed and others are newly formed.

As an implication, the dynamics is then described by \emph{two different time scales}: there is a dynamics \emph{on} the network at time scale $t$, and a dynamics \emph{of} the network at time scale $T$, and we assume that they can be \emph{separated}. 
On the shorter time scale, $t$, agents interact and in conclusion obtain a stationary value of their knowledge stock,  $x_{i}^{\mathrm{stat}}$.
On the longer time scale, $T$, the \emph{entry and exit dynamics} takes place, specifically \emph{after} the stationary solution for $x_{i}(t)$ was reached. 
That means, the interaction structure given by the network evolves on time scale $T$, whereas the knowledge stock of agents evolves on time scale $t$.

At each time step $T$ a different (quasi-)stationary value $x_{i}^{\mathrm{stat}}(T)$ is obtained. 
This can be used to distinguish between successful and not successful agents, i.e. to measure \emph{performance}.
We can \emph{rank} agents against their obtained stationary knowledge stock, $x_{i}^{\mathrm{stat}}(T)$ taken at time $T$, \emph{before} the network is changed. 
As already explained in Figure~\ref{fig:3examples}, agents without incoming links will likely have a knowledge stock of zero, as well as agents that are not part of a cycle of direct or indirect reciprocity.
Only agents that are part of collaboration cycles will reach a high value of  $x_{i}^{\mathrm{stat}}(T)$.
That means, agents well integrated into collaborations are clearly distinguishable from less integrated ones.

\paragraph{Applications. }

There are different ways to apply the above combination of nonlinear growth and entry and exit dynamics.
Using economic arguments, different nonlinear expressions can be motivated  \citep{Battiston2009}, in particular with respect to externalities, $\nu^{\mathrm{ext}}$, and saturation effects.
The impact of these assumptions on the resulting knowledge stock distribution can then be evaluated.

\begin{figure}[htbp]
  \centering
  \includegraphics[width=0.4\textwidth]{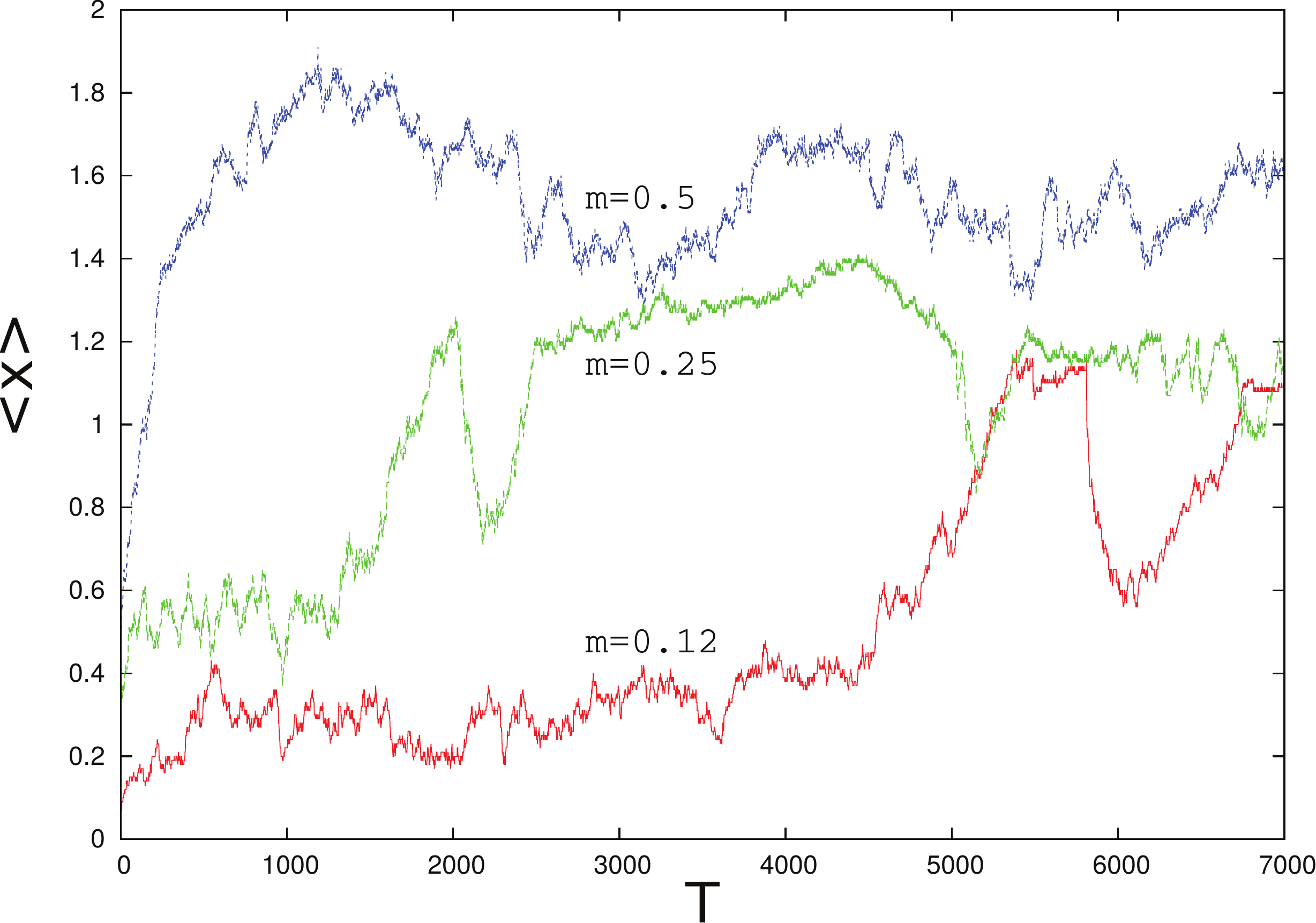}(a)\hfill
\raisebox{1ex}{\includegraphics[width=0.25\textwidth,angle=0]{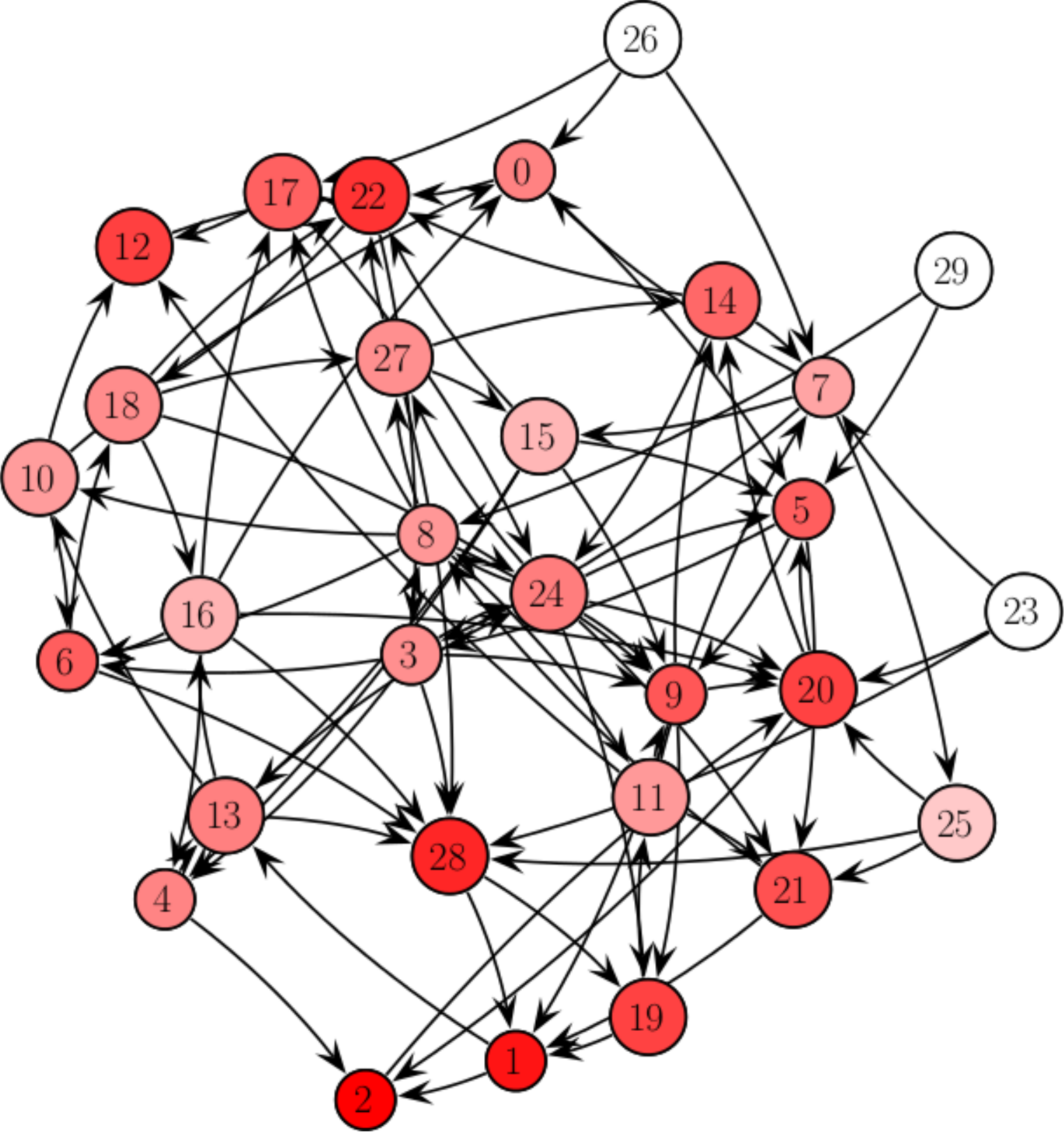}}(b) \mbox{$\quad$}
\raisebox{2ex}{\includegraphics[width=0.2\textwidth,angle=0]{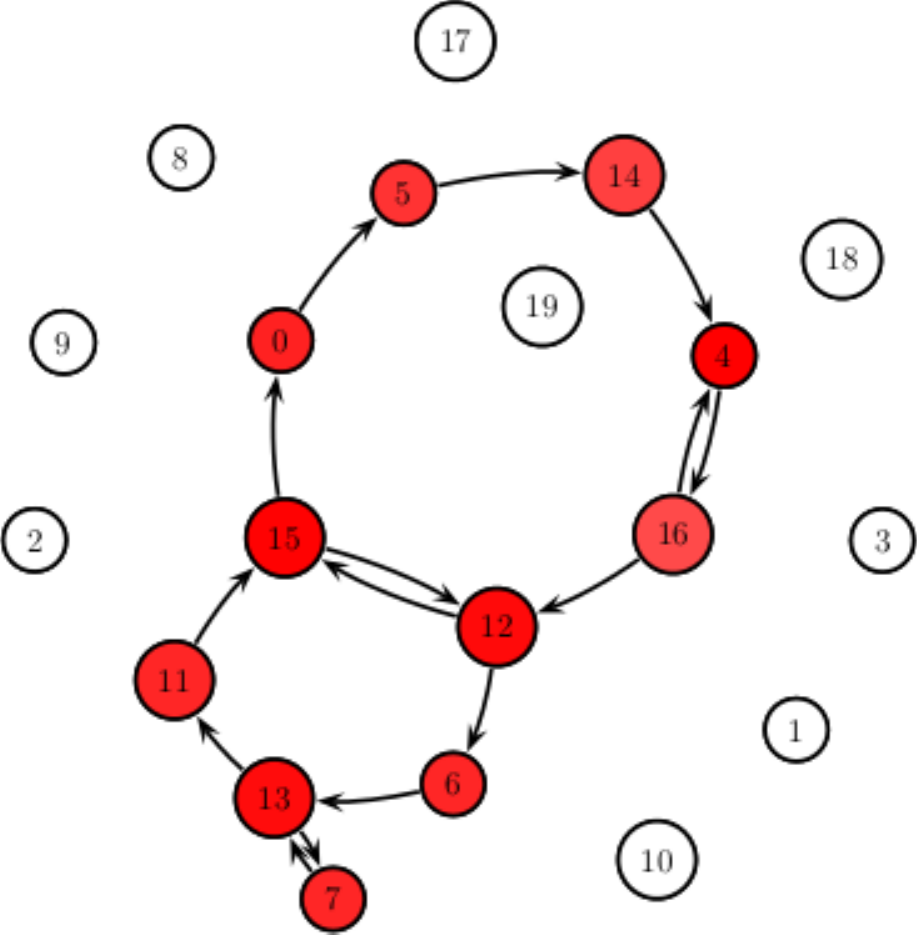}}(c) 
\caption{(a) Evolution of the average knowledge stock at time scale $T$, involving network disruptions \citep{Seufert2007}. (b,c) Different nonlinearities in Eqn.~\eqref{eq:18} combined with different mechanisms for link deletion and creation result in different network structures: (b) Extremal dynamics, no externalities. (c) Random unilateral link creation, optimal unilateral link deletion, externalities $\nu^{\mathrm{ext}}$ give higher weights to links contributing to cycles, this way fostering indirect reciprocity.
    \citep{Battiston2009}}
  \label{fig:disrupt}
\end{figure}

More important is the study of \emph{different entry and exit dynamics}. 
Its simplest form is a so-called \emph{extremal dynamics}: from the least performing agents with the lowest knowledge stock one is
randomly chosen and removed from the system together with all its collaboration links.
This agent is then replaced by a new agent, which is randomly connected to the remaining agents.
Because of the large degree of randomness involved, this type of entry and exit dynamics would describe a 
network disruption based on perturbations rather than an economic process.
Nevertheless, the dynamic outcome is quite insightful. 
Figure~\ref{fig:disrupt}(a) illustrates that the overall performance of the system, measured by means of an average knowledge stock, follows different stages:
Initially, it is very low because collaboration structures, i.e. cycles of direct and indirect reciprocity have not yet established.
After that, $\mean{x}$ constantly increases because these structures gradually improve by better integrating agents.
But if \emph{all} agents reach a high performance, the extremal dynamics will eventually destroy such structures because it removes agents from existing cylces.
This eventually leads to \emph{crashes} in the system performance which are followed by new stages of \emph{recovery}. 
This way, the system never reaches an equilibrium.

An extension of the extremal dynamics explicitly considers that entry and exit involve more than one agent. 
Instead of choosing randomly one of the agents with lowest performance, one can consider to remove a \emph{fraction} of  least performing agents \citep{mavrodiev2019} and compensate this with the entry of many more new agents.
This implies to define a threshold for the performance, which interestingly has a nontrivial impact on the overall stability of the system.
Small threshold values are able to \emph{improve} the stability by reducing the number of crashes over time \citep{mavrodiev2019}. 

\subsection{Linear growth of knowledge stock: Rational decisions}
\label{sec:costs-benefits}

The simplified entry and exit dynamics described above does not involve any decisions of agents, because they are replaced by  stochastic perturbations of the system.
In socio-economic systems, however, agents make decisions about creating or deleting links to other agents, based on their \emph{utility}.
This considers the \emph{benefits} and \emph{costs} of interactions.
To calculate these, agents need information, e.g. about the knowledge stock of their collaborators, which is not always fully available.
Therefore, decisions are based on \emph{bounded rationality}, i.e. in the absence of information also random decisions about link creation or deletion govern the dynamics.
In our case, the benefits of interactions are clearly given by the \emph{growth} of the own knowledge stock, whereas the costs result from maintaining collaborations with other agents.
The latter should be proportional to the number of links an agent has.
But here we have to consider that agents maintain \emph{outgoing} links, i.e. links that contribute to the knowledge growth of \emph{other} agents, whereas benefits depend on \emph{incoming} links from other agents (which are not necessarily the same).

This precisely describes the \emph{dilemma}: why should agents maintain links if they do not see a direct benefit from this?
\emph{Reciprocity} would be the appropriate argument for this, but only \emph{direct} reciprocity can be easily observed by an agent. To detect \emph{indirect} reciprocity would require knowledge about the interaction structure in the broader neighborhood, in case of  larger cycles even knowledge about the full system.
This becomes increasingly unrealistic.
Yet, system wide collaboration structures are empirically observed, and indirect reciprocity is an established mechanism is many social systems.

Agent-based models allow to study the conditions under which such collaboration structures \emph{emerge} despite individual utility considerations would stand against them.
We define the utility of agent $i$  as
\begin{equation}
  u_{i}(t)=B_{i}[\mathbf{A},\mathbf{x}(t)]-C_{i}[\mathbf{A},\mathbf{x}(t)]
  \label{eq:19}
\end{equation}
$\mathbf{x}(t)$ is the vector of  knowledge stock values $x_{i}(t)$ of all agents and
$\mathbf{A}$ is the adjacency matrix that describes the  current interaction structure.
Both benefits $B_{i}$ and  costs $C_{i}$ depend on these, as follows.
Benefits are assumed to be proportional to the \emph{growth} of knowledge stock, i.e. $B_{i}(t)\propto {\dot{x}_{i}(t)}/{x_{i}(t)}$ \citep{Koenig2008,Koenig2011,Battiston2012}.
The dynamics of the knowledge stock is described by Eqn.~\eqref{eq:18}, but here we drop the last two terms, i.e. we neglect saturation ($\kappa=0$) and externalities ($\nu^{\mathrm{ext}}=0$), making this a linear dynamics in $x$. 
One can then prove \citep{Koenig2008} that  $\lim_{t\to \infty}{\dot{x}_{i}(t)}/{x_{i}(t)}=\lambda^{\mathrm{PF}}(G_{i})$, where $\lambda^{\mathrm{PF}}(G_{i})$ is a property of the \emph{adjacency matrix} $\mathbf{A}$, precisely the largest real eigenvalue, also known as Perron-Frobenius (PF) eigenvalue, of the connected component $G_{i}$ agent $i$ is part of.

For the costs $C_{i}$ we consider that all outgoing links have to be maintained at a cost $c$, i.e. $C_{i}= c \sum_{j}l_{ij}= c\, k^{\mathrm{out}}_{i}$, where $k^{\mathrm{out}}_{i}$ is the \emph{out-degree} of agent $i$.
This leads to the following expression for the agent utility: 
  \begin{equation}
 u_i(t) = \frac{\dot{x}_{i}(t)}{x_i(t)} - c\, k^{\mathrm{out}}_i \quad \Rightarrow \quad u_i(T) = \lim_{t
   \to \infty} u_i(t) = \lambda^{\mathrm{PF}}(G_{i})- c\, k^{\mathrm{out}}_i
 \label{eq:20}
\end{equation}
We still have to define how agents make use of the information derived from their utility, to decide about link formation and link deletion.
We posit that these decisions are driven by \emph{utility maximization}.
If a  pair $(i,j)$ of agents is selected at random, then a  link \((i,j) \notin E(G)\), that is not part of the set of links, $E$, of the network $G$ is \emph{created} if the link $(i,j)$ increases either $u_i(T)$ or $u_j(T)$ (or both) and none decreases.
This selection scheme is known as \emph{incremental improvement}. 
As an alternative, one could also consider a random \emph{unilateral} link creation, i.e. a link to a randomly chosen agent $j$ is  already created if only $u_i(T)$ increases.

Further, if a  pair $(i,j)$ of agents is selected at random, then an existing 
link \((i,j) \in E(G)\) is \emph{deleted} if at least one of the two agents increases its utility from removing this link. 
This is known as \emph{optimal unilateral} link deletion.
An alternative could be the \emph{optimal bilateral} link deletion, which considers both agents similar to the incremental improvement scheme.

\paragraph{Applications. }
There are different ways to extent this model.
First, one could consider alternatives for \emph{calculating the utilities}, e.g. by modifying the information taken into account for the calculation, or by including nonlinear terms to reflect saturation effects, etc.
Second, one may consider alternatives for \emph{calculating the decisions} made on these utilities.
Incremental Improvement just picks the \emph{first} randomly selected pair of agents with a positive utility increase to create a link.
A different scheme would be \emph{best response}.
It creates the link only between the pair of agents that will give  the \emph{highest} increase of utility.
This would require (i) to have access to all agents and to have full information about their knowledge stocks, and (ii) to postpone decisions until all possible pairs have been considered.
It can be shown that such a scenario, even with more information, not necessarily leads to a better outcome regarding the collaboration structure.
Because of path dependent decision processes, the agent population can get trapped in \emph{suboptimal} system states \citep{Koenig2008}.

A realistic modification is to consider that link deletion involves a \emph{severance cost} \citep{Koenig2011}.
I.e. agents that have invested in establishing a collaboration will loose part of this investment when they decide to cancel the collaboration.
If these additional costs become high, agents will be more reluctant to change their collaboration structure.
Again, the system can then be trapped in suboptimal states.

\section{Conclusions}
\label{sec:conclusions}

As we have demonstrated, the law of proportionate growth is a very versatile dynamics, in particular when combined with additional dynamic assumptions.
The core dynamics simply states that the growth of a quantity $x_{i}$, i.e. $dx_{i}/dt$, is proportional to $x_{i}$.
The formal solution of this basic dynamics is exponential growth or exponential decay, dependent on the proportionality constant.
It has been observed that such a dynamics with suitable modifications can explain a larger range of empirical phenomena observed in socio-economic systems. Notably,  the ``law of proportionate growth'' was first related to the observed \emph{size distribution of firms} by R. Gibrat  in 1931 \citep{sutton1997gibrat}. 

Distributions refer to \emph{systemic} (or ``macroscopic'') properties, while the underlying dynamics is proposed for the (``microscopic'') system elements, or agents.
Hence, such agent-based models are capable of establishing the \emph{micro-macro link}, if they can explain the emergence of systemic properties from the interaction of the system elements.
Already the application by Gibrat illustrates that it needs additional assumptions to make this happen.
It is \emph{not} simply the exponential growth that reproduces the firm size distribution.
It further needs specific assumptions about the proportionality factor -- its underlying normal distribution, non-stationarity and randomness -- that only allows to obtain the correct systemic property.
Hence, what constitutes the essence of the particular phenomenon can be understood from the \emph{deviations} from the simple exponential dynamics.
And these ``deviations'' are in fact the ingredients that make a particular dynamic model an economic or a social one.
They often allow for an interpretation in a socio-economic context, as Gibrat's example witnesses.

The various applications discussed in this paper illustrate that in agent-based models the law of proportionate growth often acts at two different levels:
First, there is the growth (or the decay) $dx_{i}/dt$ proportional to the \emph{own} quantity $x_{i}(t)$, with the growth of firm sizes or individual wealth as typical examples.
Second, there is \emph{additionally} the growth $dx_{i}/dt$ proportional to the quantities $x_{j}(t)$ of \emph{other} agents, with the dynamics of opinions, reputation, or knowledge stock as examples. 
The latter requires \emph{interactions} between agents, 
for which different forms have been discussed.
There are \emph{direct} interactions between any two agents, as in the example of battling.
There are interactions \emph{restricted} by the existence of links, as in the case of networks, or thresholds, as in the bounded confidence model.
Eventually, there are also \emph{indirect} interactions resulting from the coupling to aggregated variables, for example in the wealth redistribution model or in the wisdom of crowds.

The versatility of the agent-based models discussed also results from the combination of different deterministic and stochastic growth assumptions.
There are \emph{random} proportionality factors drawn from different distributions that determine whether individual agents experience growth or decay in the long run.
Examples are the growth of firm sizes or of individual wealth. 
There are \emph{deterministic} proportionality factors, which can be either positive or negative, constant or time dependent.
Examples are saturated growth, the decay of reputation or knowledge stock.
Even more, these assumptions about growth factors can be combined in an \emph{additive} or a \emph{multiplicative} manner.
This was illustrated in the simple investment model and in the wealth redistribution model, which both combine multiplicative and additive growth. 

Despite the richness of the models that result from combining such assumptions, we still have to remark their simplicity and accessibility.
As demonstrated, in many cases we are able to formally analyze such models, to derive insights about systemic properties and critical parameter constellations.
This is in fact one of the main reasons to base our agent-based models on the law of proportionate growth as the core dynamics. 
Constructing models this way allows us to  \emph{derive expectations} about the \emph{collective} dynamics, and often to generate hypotheses, while agent-based simulations \emph{illustrate} the dynamics from an individual perspective.

\subsection*{Acknowledgements}
\label{sec:Acknowledgements}

The author thanks Nicola Perony for providing Figures~\ref{fig:comp}, \ref{fig:random}, \ref{fig:bond}, \ref{fig:battle} and Giona Casiraghi for providing Figure~\ref{fig:3examples}.

\setlength{\bibsep}{1pt}

\end{document}